\documentclass[10pt,twocolumn,letterpaper]{article}
\usepackage[T1]{fontenc}
\usepackage[pagenumbers]{cvpr}
\usepackage{tikz}
\usetikzlibrary{arrows.meta,positioning,fit,calc,backgrounds}
\usepackage{algorithm}
\usepackage{algpseudocode}
\usepackage{xr-hyper}
\newcommand{\E}{\mathbb{E}}

\newcommand{\norm}[1]{\left\lVert #1\right\rVert}
\newcommand{\ours}{VolMusCo}
\newcommand{\papertitle}{\ours{}: Learning Full-Body Motion Control with Volumetric Muscles}

\definecolor{cvprblue}{rgb}{0.21,0.49,0.74}
\usepackage[pagebackref,breaklinks,colorlinks,allcolors=cvprblue]{hyperref}
\title{\papertitle}
\author{
Wangyingqi Ding$^{1*}$\qquad
Yusen Feng$^{2*}$\qquad
Kaibo He$^{3}$\qquad
Bingchen Han$^{2}$\qquad
Libin Liu$^{2}$\\[4pt]
$^{1}$Tsinghua University\qquad $^{2}$Peking University\qquad $^{3}$Clone Incorporated\\[2pt]
{\small $^{*}$Equal contribution}
}

\newcommand{\projecturl}{https://inckieultra.github.io/VolMusCo-Project-Page/}
\newcommand{\projectlink}{%
  \ifdefempty{\projecturl}%
    {\textcolor{cvprblue}{\nolinkurl{https://[github-io-url]}}}%
    {\expandafter\url\expandafter{\projecturl}}%
}

\DeclareRobustCommand{\suppcite}[1]{%
  \begingroup
  \renewcommand{\citenumfont}[1]{S##1}%
  \cite{#1}%
  \endgroup
}

\usepackage{bibunits}

\begin{document}
\maketitle
\begin{abstract}

We present a simulation and learning framework for full-body motion control driven by volumetric muscles. Our approach establishes two-way coupling between muscles and the skeleton: muscle activations induce contraction along embedded fibers, deforming the volumetric tissue and generating forces that are transmitted to the skeleton through muscle attachments and wrapping contacts. In turn, skeletal motion changes the attachment and wrapping geometry, influencing tissue deformation and force transmission. This coupling brings volumetric deformation directly into the controlled dynamics. To make learning computationally efficient, we introduce a simplified tetrahedral muscle representation that preserves essential volumetric deformation and force transmission while substantially reducing simulation cost. Building on this representation, we develop a model-based reinforcement learning framework that learns a differentiable world model from physical transitions and uses predicted future motion to improve the control policy. On LaFAN1 motion tracking, our method achieves a tracking success rate of 97.9\% and a mean local-body position error of 5.52\,cm within 10\,h of training. Project page and video demonstrations are available at \projectlink.
\end{abstract}

\section{Introduction}

\label{sec:intro}

Human motion emerges from the coordinated interaction between skeletal dynamics and muscle contraction. Muscle deformation and skeletal motion are mutually dependent and evolve through a coupled physical process. Neural activations induce contraction along muscle fibers, deforming the surrounding tissue and generating forces that drive the skeleton, while also shaping the body's appearance. Conversely, skeletal motion changes muscle attachment and wrapping configurations, which in turn influence tissue deformation and force transmission. 

Existing muscle-driven character control predominantly relies on Hill-type muscle models, which represent the character's musculature as hundreds of one-dimensional, massless actuators attached to rigid skeletal segments~\cite{hill,zajac,millard}.
Recent methods have advanced motion control with these models through hierarchical coordination~\cite{mass}, muscle-feasible motion tracking~\cite{magnet}, and GPU-parallel imitation learning~\cite{musclemimic}. However, this simplified representation cannot explicitly capture volumetric tissue deformation, such as transverse bulging, or interactions between muscles and surrounding anatomical structures. Accounting for dynamic effects of muscle motion, such as inertia, also requires additional modeling~\cite{muscleinertia,inertialmusculotendons}.

Volumetric muscle models provide the spatial degrees of freedom needed to capture these deformations and interactions. They have been widely used in computer graphics to simulate detailed muscle deformation and anatomical interactions~\cite{mcadams2011,teran,lee2011comprehensive}. In many such applications, muscle deformation is computed in response to prescribed skeletal motion, forming a one-way coupling from the skeleton to the muscle tissue. Recent work has explored learning motion control using one-dimensional muscle models and applying the resulting muscle activations to volumetric simulations~\cite{rlmuscle}. However, the coupling remains one-way, as volumetric simulation is outside the control loop and tissue responses do not affect skeletal motion during control. Prior work has also established two-way coupling between volumetric muscles and skeletal dynamics~\cite{activevol,volcontrol}, allowing muscle deformation and its mechanical reactions to jointly determine skeletal motion. However, the high dimensionality of the coupled system makes learning-based control challenging, and existing approaches have therefore focused on more specialized settings, such as upper-body manipulation or task-specific trajectory optimization, rather than scalable full-body motion control.

In this paper, we present \ours{}, a simulation and learning framework for full-body motion control driven by volumetric muscles. Our approach establishes two-way coupling between muscle tissue and skeletal dynamics, integrating tissue deformation and its mechanical effects directly into the control loop.

A major challenge in learning under such coupled dynamics is the computational cost of repeatedly solving volumetric tissue equilibrium during reinforcement learning. To make physical simulation efficient, we introduce a simplified tetrahedral muscle representation that preserves essential volumetric deformation and fiber-based actuation while substantially reducing the cost of tissue simulation. We also adopt a model-based reinforcement learning framework that learns a differentiable world model from physical transitions collected in the volumetric simulator. The learned model predicts future physical states under muscle excitations, allowing the policy and critic to improve tracking behavior through short predicted trajectories without repeatedly executing expensive tissue solves.

We evaluate \ours{} on full-body motion tracking using LaFAN1 dataset~\cite{lafan} and additional fitness and object-manipulation motions. On LaFAN1 clips, our method achieves a tracking success rate of 97.9\% and a mean local-body position error of 5.52\,cm within 10 hours of training. We also conduct controlled experiments to evaluate the fidelity of the simplified muscle representation, the effect of deformation-dependent force feedback, and the learning efficiency gained through world-model prediction. The paper makes two contributions:

\begin{itemize}[leftmargin=*,itemsep=2pt,topsep=3pt]

\item We present a new learning framework for full-body motion control driven by two-way coupled volumetric muscles. To our knowledge, this is the first framework to enable such control.

\item We introduce a computationally efficient design that combines a simplified tetrahedral muscle representation with model-based reinforcement learning, enabling scalable learning across a diverse range of full-body motions.

\end{itemize}

\section{Related Work}
\label{sec:related}

\paragraph{Learning muscle-driven control.}
Early controllers combine muscle reflexes with legged mechanics~\cite{geyer}, optimize biologically motivated locomotion objectives~\cite{bioloco}, and adapt muscle control across biped morphologies~\cite{flexible}. Many-muscle locomotion~\cite{manymuscle} extends control to detailed humanoid musculature. OpenSim~\cite{opensim} and detailed musculoskeletal models~\cite{rajagopal} support internal-force analysis. OpenSim Moco~\cite{moco} enables muscle-driven motion through trajectory optimization. MASS~\cite{mass} instead learns skeletal coordination and muscle recruitment at separate levels. Generative GaitNet~\cite{gaitnet} conditions control on anatomy and gait requirements; Bidirectional GaitNet~\cite{bidirectional} also estimates anatomical conditions from gait. MAGNET~\cite{magnet} addresses diverse-motion tracking through muscle-feasible torque clipping, a muscle coordination network, and staged adaptation. Other approaches address overactuation through structured exploration, as in DEP-RL~\cite{deprl}, or state-dependent synergistic action representations, as in DynSyn~\cite{dynsyn}. MyoSuite~\cite{myosuite} supplies contact-rich musculoskeletal tasks. KINESIS~\cite{kinesis} learns muscle-driven locomotion through imitation, while MuscleMimic~\cite{musclemimic} combines full-body assets, retargeting, and GPU-parallel imitation learning. MS-Emulator~\cite{wei2026msemulator} learns muscle-driven imitation with adversarial rewards and a value-guided flow policy. These systems broaden motion coverage, but kinematics alone leave muscle recruitment ambiguous. Muscles in Action~\cite{mia} and Muscles in Time~\cite{mint} study muscle--motion relationships through measured or simulated labels. MUSIC~\cite{music} distills high-frequency tracking for dexterous piano performance into a low-dimensional controller reusable by a slower music-conditioned policy. Wang et al.~\cite{elite} train direct excitation policies for athletic locomotion using task rewards and high-throughput simulation, without motion demonstrations. MuscleVAE~\cite{musclevae} is the closest model-based muscle-control precedent, incorporating muscle-length control, fatigue, and a generative skill representation. We reuse the model-based control principle without its generative VAE; our focus is deformation-dependent force feedback to the skeleton.

\paragraph{World models for motion tracking.}
DeepMimic~\cite{deepmimic} establishes reference-guided reinforcement learning for simulated skills. PULSE~\cite{pulse} learns reusable humanoid motion representations, and MaskedMimic~\cite{maskedmimic} formulates control from partial motion descriptions as motion inpainting. BeyondMimic~\cite{beyondmimic} combines agile tracking with guided-diffusion control; SONIC~\cite{sonic} scales a generalist tracker through model capacity, motion data, and compute. Learned dynamics provide another route to reusing experience. Dyna~\cite{dyna} integrates environment interaction and model-based updates, while model-based policy optimization~\cite{mbpo} uses short model rollouts branched from real data to limit model bias. Dreamer~\cite{dreamerv3} learns behavior using future states predicted in latent space. For physical animation, SuperTrack~\cite{supertrack} optimizes tracking through learned transition derivatives, ControlVAE~\cite{controlvae} jointly learns dynamics and generative control, and MoConVQ~\cite{moconvq} learns scalable discrete control representations. SuperMimic~\cite{supermimic} trains a shared world model and multi-motion actor together, with a critic extending short model-predicted trajectories. MorphMimic~\cite{morphmimic} studies sharing across humanoid morphologies.
We follow the same model-based learning paradigm and extend it to motion control with volumetric muscles.
SHAC~\cite{shac} and AHAC~\cite{ahac} additionally motivate our attention to finite-horizon gradients and stiff contact dynamics.

\paragraph{Volumetric muscle simulation and control.}
Teran et al.~\cite{teran} construct and simulate anatomical skeletal-muscle geometry. Active Volumetric Musculoskeletal Systems~\cite{activevol} couples deformable muscles in close contact to multibody dynamics. Sueda et al.~\cite{sueda} and Sachdeva et al.~\cite{sachdeva} address tendon routing and transmission in detailed hand models. Lee et al.~\cite{volcontrol} integrate volumetric actuation into dexterous motion control and trajectory optimization. These works provide direct precedents for tissue-to-bone force feedback. Projective dynamics~\cite{projectivedynamics} and XPBD~\cite{xpbd} provide established deformation solvers. Romeo et al.~\cite{musclefascia} develop anisotropic muscle-fiber and superficial-fascia constraints within XPBD for visual effects animation; MGPBD~\cite{mgpbd} uses algebraic multigrid to accelerate a global XPBD solve. RLMuscle~\cite{rlmuscle} instead uses geometric multigrid for volumetric muscles and learns activations on a separate one-dimensional surrogate. Its measured EMG prior helps select among redundant activation solutions, while the three-dimensional stage produces the resulting deformation. Our formulation studies model-based RL with quasi-static deformable bellies, simplified active-fiber energy, and endpoint/wrapping reactions inside skeletal dynamics. Bone pose changes the attachment and contact energies; their reactions supply skeletal loads. Virtual-work and energy-gradient tests assess numerical consistency alongside learned tracking.

\section{Volumetric Muscle Control}
\label{sec:method}

\begin{figure}[t]
  \centering
  \input{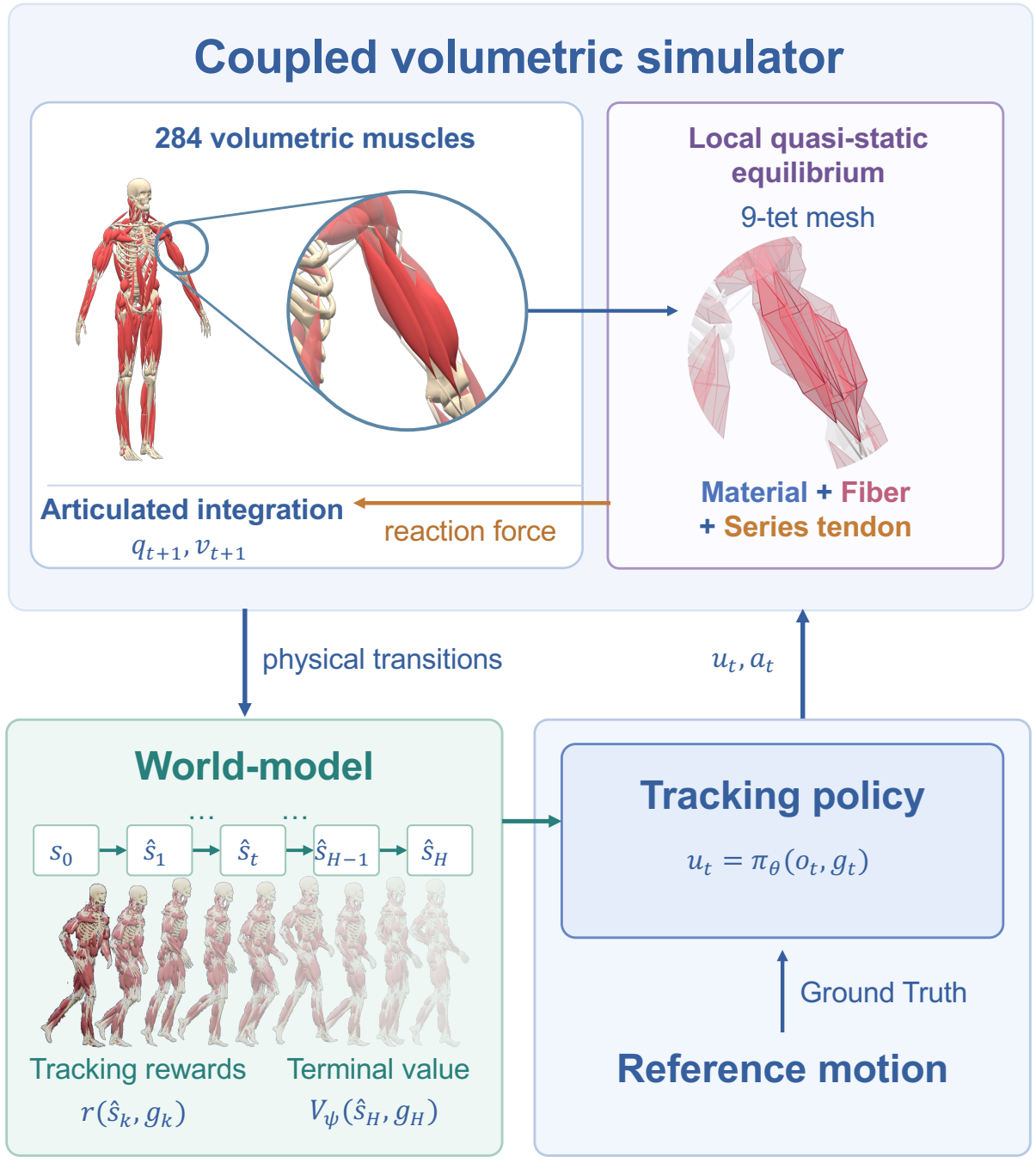}
  \caption{\textbf{Volumetric muscle control.} Skeletal pose changes muscle attachment and wrapping geometry; quasi-static tissue equilibrium generates forces that drive the skeleton. Physical simulator transitions train the world model; its predictions are used to improve the tracking policy.}
  \label{fig:overview}
\end{figure}

We learn motion tracking with volumetric muscles using the pipeline in Figure~\ref{fig:overview}. A policy controls muscle activation to generate forces that drive skeletal motion. Skeletal pose $q$ changes attachment and wrapping energies, altering muscle equilibrium and the resulting forces. We train a differentiable world model on these transitions and use its predictions of future motion to improve the tracking policy.

\subsection{Motion-tracking objective}
\label{sec:problem}

The reference motion is the tracking target, represented by $g_t$ at control step $t$. The policy $u_t=\pi_\theta(O(s_t),g_t)$ maps the physical-state observation $O(s_t)$ and reference input $g_t$ to muscle excitation vector $u_t$. Here $s_t$ is the simulator state vector, $O$ is the observation function, and $\theta$ denotes policy parameters. We optimize
\begin{equation}
\begin{aligned}
 \max_\theta\;&\E_{\pi_\theta,\mathcal F_{\rm vol}}
 \left[\sum_{t=0}^{T-1}\gamma^t r(s_t,g_t)\right],\\
 &s_{t+1}=\mathcal F_{\rm vol}(s_t,u_t).
\end{aligned}
\label{eq:rlobjective}
\end{equation}
Here $t$ indexes control steps, $T$ is the rollout length, $\gamma$ is the discount factor, and $r$ measures agreement with the reference. $\mathcal F_{\rm vol}$ is the volumetric muscle simulation engine. State $s_t$ includes skeletal positions $q_t$, velocities $v_t$, and muscle activations $a_t$.

\subsection{Model-based learning of motion control}
\label{sec:worldmodel}

We follow the model-based RL paradigm~\cite{supertrack,controlvae,musclevae,supermimic}, and extend it to motion control with volumetric muscles.
We run the policy in the volumetric muscle simulator and store the resulting state transitions in a replay buffer. The world model predicts the next state from the current state and muscle excitation:
\begin{equation}
 \hat s_{j+1}=f_\omega(\hat s_j,u_j),
\label{eq:worldmodel}
\end{equation}
where $f_\omega$ is the transition model with parameters $\omega$, $\hat s_j$ is the predicted state at rollout step $j$, and $u_j$ is the muscle excitation vector. Initial state $\hat s_0$ comes from physical replay. Model fitting unrolls recorded commands and minimizes
\begin{equation}
 \mathcal L_{\rm WM}=\E_{\mathcal B}\sum_{j=1}^{K}
 \norm{\hat s_j-s_j}_W^2.
\label{eq:wmloss}
\end{equation}
The replay buffer $\mathcal B$ supplies sequences of $K$ transitions; $s_j$ is the recorded state vector, and positive-semidefinite matrix $W$ weights the squared state-prediction errors. To improve the policy, we predict future states with the world model using commands from the current policy and maximize
\begin{equation}
 J(\theta)=\E\Big[\sum_{j=0}^{H-1}\gamma^j r(\hat s_j,g_j)+\gamma^H V_\psi(\hat s_H,g_H)\Big],
\label{eq:training}
\end{equation}
where $H$ is the prediction horizon, $\theta$ is the policy parameter vector, and critic $V_\psi$, with parameters $\psi$, predicts terminal tracking return. Reward $r$, reference input $g_j$, and discount $\gamma$ retain their tracking meanings. The expectation averages trajectories predicted from initial states sampled from the replay buffer.

Training alternates between fitting the world model to simulator transitions and updating the policy through the fitted world model.

\subsection{Two-way volumetric muscle dynamics}

Muscle forces drive skeletal motion, while skeletal motion changes muscle energy, deformation, and force transmission. We update muscle attachments and wrapping geometry from skeletal pose $q$. Policy commands change muscle activation and thus the active contraction energy within the tissue. For the current pose and activations, we solve for the equilibrium tissue shape and apply the resulting attachment and contact forces to the bones, completing the two-way coupling.

\paragraph{Muscle equilibrium.}
\label{sec:deformation}
We represent each muscle in our human model as a deformable tetrahedral mesh. At each muscle update time $t$, we hold skeletal pose and activation fixed and find the vertex positions that minimize the muscle energy. For brevity, we omit the time subscript $t$ and the muscle index on individual energy components below:
\begin{equation}
\begin{aligned}
 x_m^*&\approx\arg\min_{x_m}E_m(x_m;q,a_m),\\
 E_m&=E_{\rm mat}+E_{\rm act}+E_{\rm att}\\
 &\quad+E_{\rm con}+E_{\rm bar},\qquad E=\sum_m E_m,
\end{aligned}
\label{eq:potential}
\end{equation}
where $m$ indexes muscles, $x_m$ collects the world-space positions of the vertices in muscle $m$'s mesh, $x_m^*$ gives their approximate equilibrium positions, and $a_m$ is muscle activation. $E_{\rm mat}$ is the passive material energy, $E_{\rm act}$ the active contraction energy, $E_{\rm att}$ the endpoint attachment energy, $E_{\rm con}$ the wrapping contact energy, and $E_{\rm bar}$ a volume barrier against element collapse. The total energy $E$ is the sum of the per-muscle energies $E_m$.

\paragraph{Passive and active energies.}
For each tetrahedron, we compute the deformation gradient
\begin{equation}
 F_e=D_{s,e}D_{0,e}^{-1},\qquad J_e=\det F_e,
\label{eq:deformationgradient}
\end{equation}
where $e$ indexes tetrahedra, and $D_{s,e}$ and $D_{0,e}$ contain three edge vectors in the deformed and rest configurations. Thus $F_e$ describes local deformation, while $J_e$ is the deformed-to-rest volume ratio. The passive energy is
\begin{equation}
 E_{\rm mat}=\sum_e\frac{w_e}{2}
 \left(\|F_e-R_e\|_F^2+\nu\|F_e-Z_e\|_F^2\right),
\label{eq:material}
\end{equation}
where $w_e$ is a material weight incorporating the tetrahedron's rest volume, $\nu$ controls the volume penalty relative to the shape penalty, and $\|\cdot\|_F$ is the Frobenius norm. The rotation $R_e$ and volume-preserving projection $Z_e$ are computed from $F_e$ using projective dynamics (PD)~\cite{projectivedynamics,volcontrol}. The two energy terms penalize shape distortion and volume change, respectively.

We use a quadratic fiber energy for active contraction~\cite{volcontrol}:
\begin{equation}
 E_{\rm act}=\sum_e\frac{a_m c_e}{2}\|F_e d_e\|^2.
\label{eq:fiberprofile}
\end{equation}
Here $a_m$ is the activation of muscle $m$, $d_e$ is the unit rest-fiber direction, and $c_e$ is active stiffness times rest volume. The length $\|F_ed_e\|$ is fiber stretch. Increasing activation increases the tendency to shorten fibers; the achieved shape balances this contraction against material, attachment, and contact forces. The volume barrier $E_{\rm bar}$ discourages flattened or inverted tetrahedra during the equilibrium solve.

\begin{figure}[t]
\centering
\includegraphics[width=\linewidth]{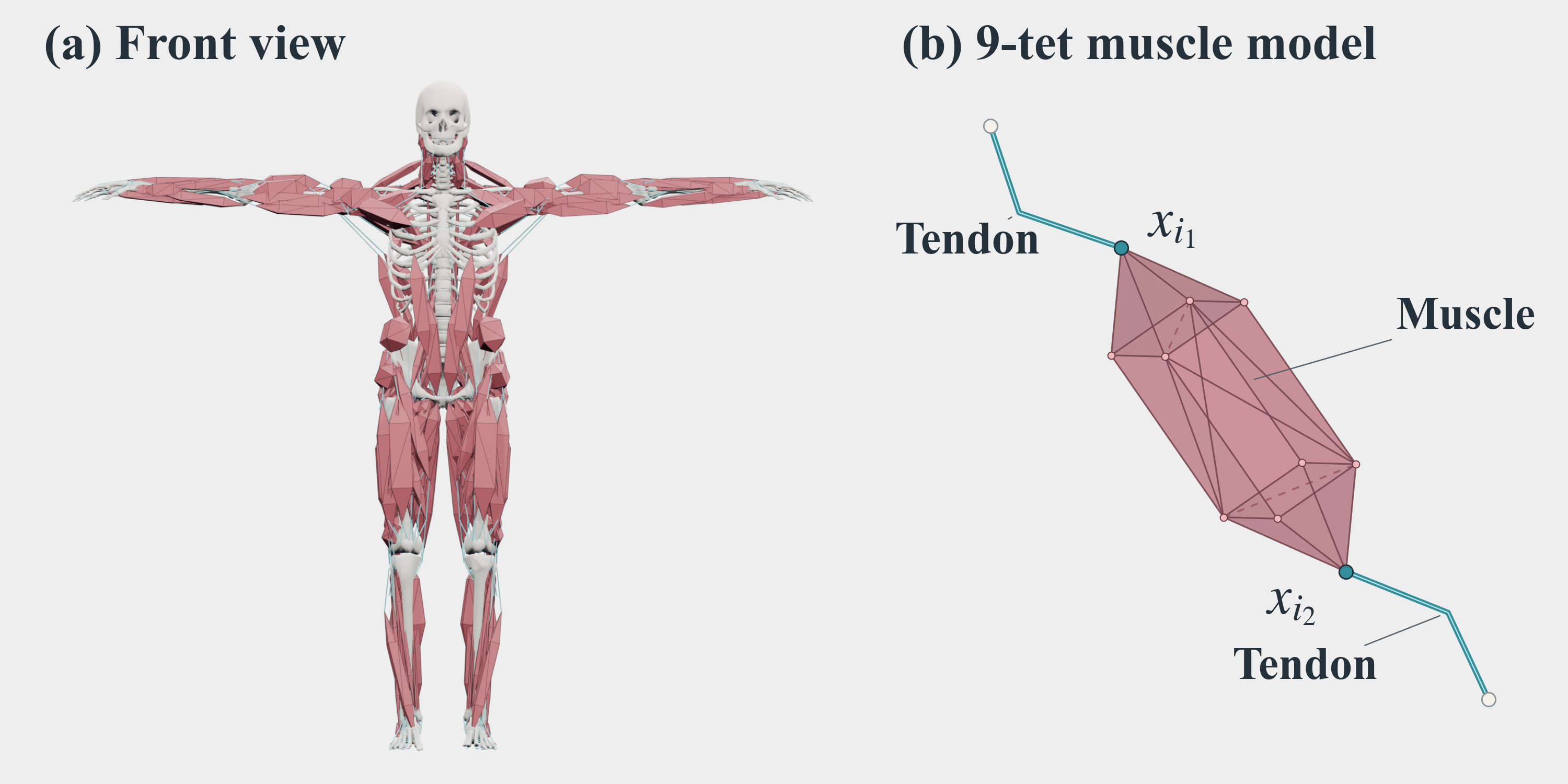}
\caption{\textbf{Coarse full-body muscle geometry.} (a) Full-body geometry comprising 284 muscles and 2,556 tetrahedra.
(b) A 9-tet muscle mesh with 10 vertices and tendon connections at the tips $\mathbf{x}_{i_1}$ and $\mathbf{x}_{i_2}$.}
\label{fig:asset}
\end{figure}

\begin{figure}[t]
    \centering
    \includegraphics[width=\linewidth]{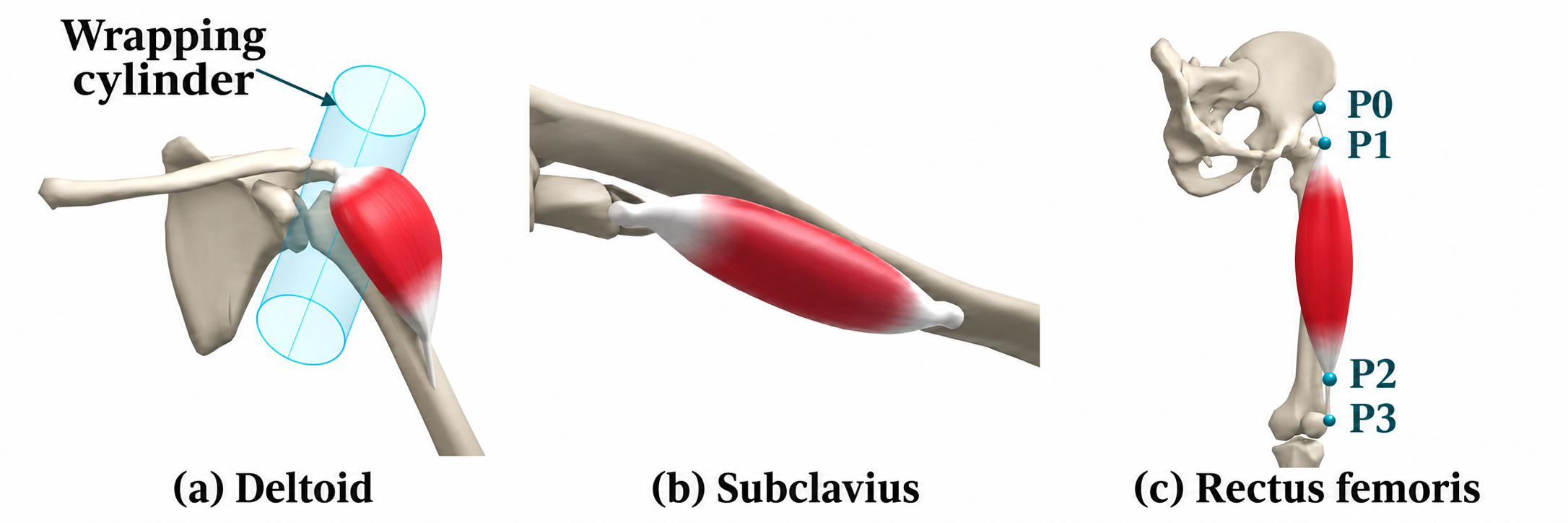}
    \caption{\textbf{Muscle--joint interactions.} We use 3 modeling configurations according to how muscle bellies and tendons interact with joints.
(a) The deltoid belly wraps around the shoulder joint; we model its wrapping contact with a cylinder.
(b) For the subclavius, belly wrapping has little effect on joint motion, so we use direct bone attachment.
(c) The rectus femoris has long, slender tendons whose paths determine how muscle forces act on the joints; we represent these routes with bone-fixed waypoints labeled $P_i$.}
    \label{fig:muscle-wrapping}
\end{figure}

\paragraph{Muscle attachments.}
Skeletal pose $q$ enters the muscle energy through the endpoint connections and wrapping geometry illustrated in Figure~\ref{fig:muscle-wrapping}. For endpoint $h$, routed-tendon energy $E_{{\rm route},h}$ models a tendon following bone-fixed waypoints; direct-attachment energy $E_{{\rm direct},h}$ models a spring to a bone:
\begin{equation}
\begin{aligned}
 L_h&=\|x_{i_h}-y_{h,1}(q)\|\\
 &\quad+\sum_{j=1}^{n_h-1}\|y_{h,j+1}(q)-y_{h,j}(q)\|,\\
 E_{{\rm route},h}&=\tfrac12 k_h[L_h-L_{0,h}]_+^2,\\
 E_{{\rm direct},h}&=\tfrac12 k_h\|x_{i_h}-y_h(q)\|^2.
\end{aligned}
\label{eq:seriestendon}
\end{equation}
Here $i_h$ identifies the muscle-tip vertex $x_{i_h}$ shown in Figure~\ref{fig:asset}(b), and $y_{h,j}(q)$ are $n_h$ ordered bone-fixed waypoints. Length $L_h$ includes the tip-to-first-waypoint segment and the remaining route; $L_{0,h}$ is slack length, $k_h$ is connection stiffness, and $[z]_+=\max(z,0)$ is the positive part. Point $y_h(q)$ is the direct bone attachment. The routed tendon pulls only when taut; the direct spring is bilateral. Each end selects one connection type, giving
\begin{equation}
 E_{\rm att}=\sum_{h\in\mathcal R_m}E_{{\rm route},h}
 +\sum_{h\in\mathcal D_m}E_{{\rm direct},h}.
\label{eq:attachment}
\end{equation}
Here $\mathcal R_m$ and $\mathcal D_m$ contain the routed and directly attached endpoints of muscle $m$, respectively. These disjoint sets partition its two endpoints, so each contributes through its configured connection.

\paragraph{Wrapping contact.}
In addition to endpoint connections, muscle bellies interact with bone-carried cylinders that guide wrapping around joints. The contact energy is
\begin{equation}
 E_{\rm con}=\sum_{s\in\mathcal S}\tfrac12 k_c[R_s-\rho_s(x,q)]_+^2,
\label{eq:contact}
\end{equation}
where $x$ contains tissue vertices, $q$ is skeletal pose, and $\mathcal S$ is the set of checked belly-edge and endpoint-connection segments. Parameter $k_c$ is contact stiffness, $R_s$ is the cylinder radius including contact clearance, and $\rho_s$ is the minimum radial distance between segment $s$ and its cylinder axis. Both bone-side segment endpoints and cylinder poses depend on $q$. The contact penalty produces equal and opposite reactions on tissue and bones.

\paragraph{Muscle solver.}
We solve the equilibrium problem in Eq.~\ref{eq:potential} using PD local/global updates of the vertex positions, with updated attachment/contact terms and line search. We monitor convergence using nodal force residuals:
\begin{equation}
 f_{m,i}=-\frac{\partial E_m}{\partial x_{m,i}},\qquad
 r_m=\max_i\|f_{m,i}\|_2\leq\varepsilon.
\label{eq:equilibriumresidual}
\end{equation}
Here $i$ indexes muscle vertices, $f_{m,i}$ is nodal force, $r_m$ is the largest nodal residual, and $\varepsilon=10^{-3}F_0$ is the stopping threshold. The force scale is $F_0=\max_m F_{\max,m}$, where $F_{\max,m}$ denotes the maximum force of muscle $m$.

\paragraph{Skeletal forces.}
\label{sec:force}
At the computed equilibrium, we assemble the attachment and contact reactions into a generalized skeletal force:
\begin{equation}
\begin{aligned}
 Q^{\rm mus}&=-\left.\frac{\partial E(x;q,a)}{\partial q}\right|_{x=x^*}\\
 &=\sum_b\left(J_{v,b}^{\top}F_b+J_{\omega,b}^{\top}\tau_b\right),
\end{aligned}
\label{eq:generalized}
\end{equation}
where $x^*$ collects the muscle equilibria, $a$ collects activations, and $Q^{\rm mus}$ is generalized muscle force. The derivative holds tissue positions and activation fixed and denotes skeletal virtual work. For bone $b$, $F_b$ and $\tau_b$ are resultant attachment/contact force and torque about its center of mass. Jacobians $J_{v,b}$ and $J_{\omega,b}$ map skeletal velocity to that body's translational and angular velocities.

\paragraph{Skeletal integration.}
We advance the skeleton with the computed muscle load, holding it fixed between muscle updates:
\begin{equation}
 (q_{n+1},v_{n+1})=\mathcal D_{\Delta t}(q_n,v_n;Q_k^{\rm mus}).
\label{eq:rigidstep}
\end{equation}
Here $n$ indexes rigid steps, $k$ indexes the enclosing muscle-update interval, and $Q_k^{\rm mus}$ is held throughout that interval. The map $\mathcal D_{\Delta t}$ advances skeletal positions $q_n$ and velocities $v_n$ by rigid step size $\Delta t$, including inertia, gravity, and external contacts. The next muscle solve uses the updated bone pose, closing the two-way loop.

\section{Experiments}

\label{sec:experiments}

We evaluate \ours{} on a large motion set, measuring motion-tracking accuracy, mechanical consistency, and computational efficiency. The experiments separate actuation-model choice, deformation-dependent force feedback, mesh resolution, and world-model learning efficiency.

\subsection{Assets and training configurations}

We construct the high-resolution muscle meshes in Figure~\ref{fig:muscle-wrapping} by sweeping cross-sections along MASS Hill-type muscle paths~\cite{mass} and fitting their profiles and target volumes to RLMuscle anatomical meshes~\cite{rlmuscle}.

For RL, we fit the 10-vertex, 9-tet mesh shown in Figure~\ref{fig:asset}(b) to each high-resolution muscle belly. We select 2 points on the Hill path to define the belly endpoints and constrain the 2 tip vertices to these positions during mesh construction. The remaining 8 vertices form 2 quadrilateral cross-sections, whose positions are optimized to match the target volume and approximate the reference belly shape while retaining the 9-tet connectivity. This fitting accommodates the curved or flattened profiles of selected shoulder and gluteal bellies. Axial reference fibers and the tendon paths outside the belly are retained. The model contains 284 muscle units, 2,840 vertices, and 2,556 tetrahedra.

The motion references combine 60 LaFAN1~\cite{lafan} clips from 5 subjects, custom keyframed lifting exercises and object-manipulation sequences, and motions reconstructed from monocular videos. We use the body-landmark estimator from BlazePose GHUM Holistic~\cite{mediapipepose}, implemented as MediaPipe Pose Full, to estimate 2D image landmarks and pelvis-relative 3D landmarks. We adapt these landmarks to our skeleton's bone lengths, recover joint angles with joint-constrained inverse kinematics, and apply temporal smoothing.

We use MuJoCo (MJWarp) as the simulation engine and solve tissue equilibrium in Warp, running 2,048 parallel environments on a single NVIDIA RTX 4090 GPU (24\,GB). We calibrate material and connection parameters per muscle. The 3 muscle--joint configurations in Figure~\ref{fig:muscle-wrapping} are wrapping contact with bone-carried cylinders (a), direct bone attachment (b), and tendon routing through bone-fixed waypoints (c).

\subsection{Local solve and multi-rate collection}

\label{sec:multi_rate}

We run the controller at 50\,Hz, solve muscle equilibrium at 100\,Hz, and integrate skeletal dynamics at 1\,kHz. Each 20\,ms control interval contains 2 muscle solves and 20 rigid steps. We hold the generalized muscle load for 10 rigid steps after each solve and use the updated skeletal pose for the next solve.

PD uses local/global updates with line search and a maximum nodal force residual threshold $\varepsilon=10^{-3}F_0$, where $F_0=\max_m F_{\max,m}$ and $F_{\max,m}$ is the maximum force of muscle $m$. The volume barrier enforces a lower threshold $J_{\min}=0.1$ on the local deformed-to-rest volume ratio $J_e=\det F_e$. Muscle and wrapping solves are capped at 600 and 50 iterations, respectively. For each of 240 independently initialized poses sampled from 60 LaFAN1 clips, we solve all 284 muscles at 3 activation levels: 0.005, 0.5, and 1.0. All $3\times240\times284=204{,}480$ muscle solves satisfy the residual threshold. We reject a candidate pose if any muscle solve fails to converge.

\subsection{Main results}
\label{sec:mainresults}

\begin{figure*}[t]
\centering
\input{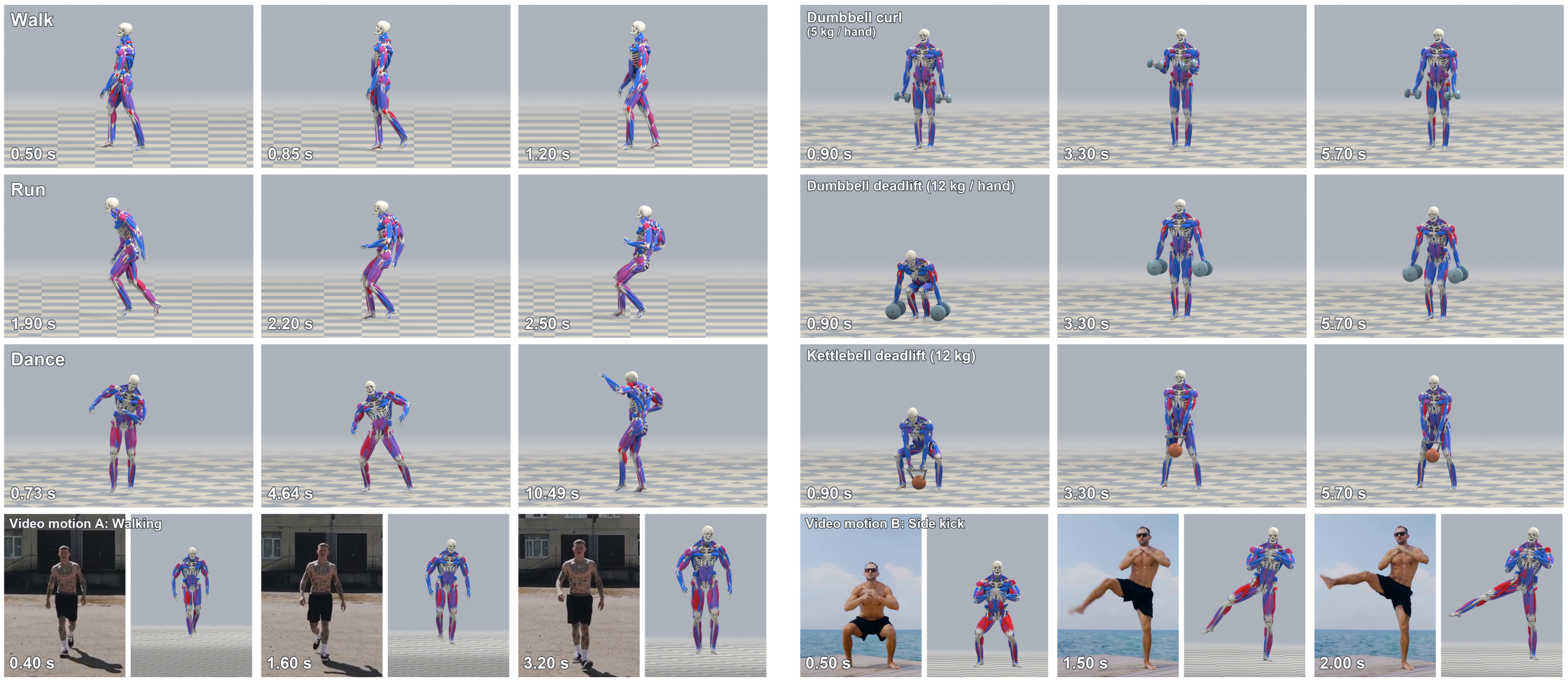}
\caption{\textbf{Volumetric motion tracking.} The rows show a LaFAN1 motion, a custom fitness motion, and a motion reconstructed from a monocular video. Muscle color indicates activation level, ranging from blue (zero activation) to red (full activation).}
\label{fig:motiongallery}
\end{figure*}

On LaFAN1 clips, \ours{} achieves a tracking success rate of 97.9\% and a mean local-body position error of 5.52\,cm within 10\,h of training. The simulator collects 815 physical control transitions per second in aggregate, with a total training cost of 16 GPU-hours. Figure~\ref{fig:motiongallery} presents tracking examples from LaFAN1, custom fitness motions, and motions reconstructed from monocular videos. Each row follows a single closed-loop rollout at aligned timestamps. Cameras and activation color scales are fixed within each sequence. Video demonstrations and interactive visualizations are available at \projectlink.

\subsection{Mesh-resolution ablation}

\label{sec:mesh_ablation}

We compare the high-resolution 230-tet muscle meshes illustrated in Figure~\ref{fig:muscle-wrapping} with the 9-tet meshes in Figure~\ref{fig:asset}(b) to evaluate simulation fidelity and RL collection cost. Both resolutions use the same material/active energies, endpoint connections, wrapping geometry, rest volume, axial length, fiber directions, physical parameters, and stopping targets.

Table~\ref{tab:mesh_ablation} reports relative tension discrepancy, scripted joint-motion discrepancy, and aggregate physical control transitions per second at a fixed environment count. We use matched activation/path sweeps and short free-joint rollouts to measure simulation fidelity. Timing includes tissue solves, attachment/contact reaction assembly, and rigid stepping.

\begin{table}[t]
\centering\small
\caption{\textbf{Mesh-resolution ablation.}
We compare 9-tet and 230-tet muscle meshes.
$\Delta_T$ is the maximum relative $L_2$ discrepancy
in tendon tensions over 4 static conditions, normalized by the 9-tet result; direct-attachment and wrapping-contact forces are excluded. $\Delta_q$ is the pooled joint-angle RMSE over 50 control timestamps and 50 scalar joints in a 1\,s full-body rollout, excluding the free root. Samples/s reports physical control transitions per second for the coarse and fine meshes, respectively.}
\label{tab:mesh_ablation}
\begin{tabular}{@{}lcccc@{}}
\toprule
Mesh pair & Tets/unit & Samples/s & $\Delta_T$ (\%) &
$\Delta_q$ ($^\circ$)\\
\midrule
Coarse / fine & 9 / 230 & 815 / 0.05 & 9.46 & 1.87\\
\bottomrule
\end{tabular}
\end{table}

\subsection{Force transmission and mechanical checks}

\label{sec:forceexperiments}

We sweep activation and skeletal pose, recording fiber stretch, routed-tendon tension, direct-attachment forces, contact penetration, and generalized torque. Fixed-shape finite differences of attachment/contact energy check the reaction map. Slack-load tests disable direct attachments and contact, isolating the routed tendon; mixed-connection tests include every reaction channel. The maximum isolated slack-tendon load is $0$\,N and generalized-load error is $0.0441\%$ (rotational) and $1.31\%$ (translational).

We assess the complete reaction-to-generalized-force map through virtual power. A separate reduced-energy test resolves local equilibrium after each pose perturbation. The relative virtual-power discrepancy is $8.87\times10^{-7}\%$ and the free-motion energy drift is $23.1\%$.

\subsection{Deformation-feedback ablation}

To isolate deformation-dependent force feedback, we compare \ours{} with a frozen nominal map $\widetilde Q^{\rm mus}(q,a)$. We fit a separate MLP with 2 hidden layers of width 64 for each muscle to equilibrium reactions from separate pose/activation samples at nominal material parameters. The networks predict attachment, tendon, and wrapping reactions from current geometry and activation, which are assembled into skeletal loads. Network weights and nominal material response remain fixed, while inputs and loads update with skeletal motion; no tissue equilibrium solve is performed for the map.

\begin{figure}[t]
\centering
\input{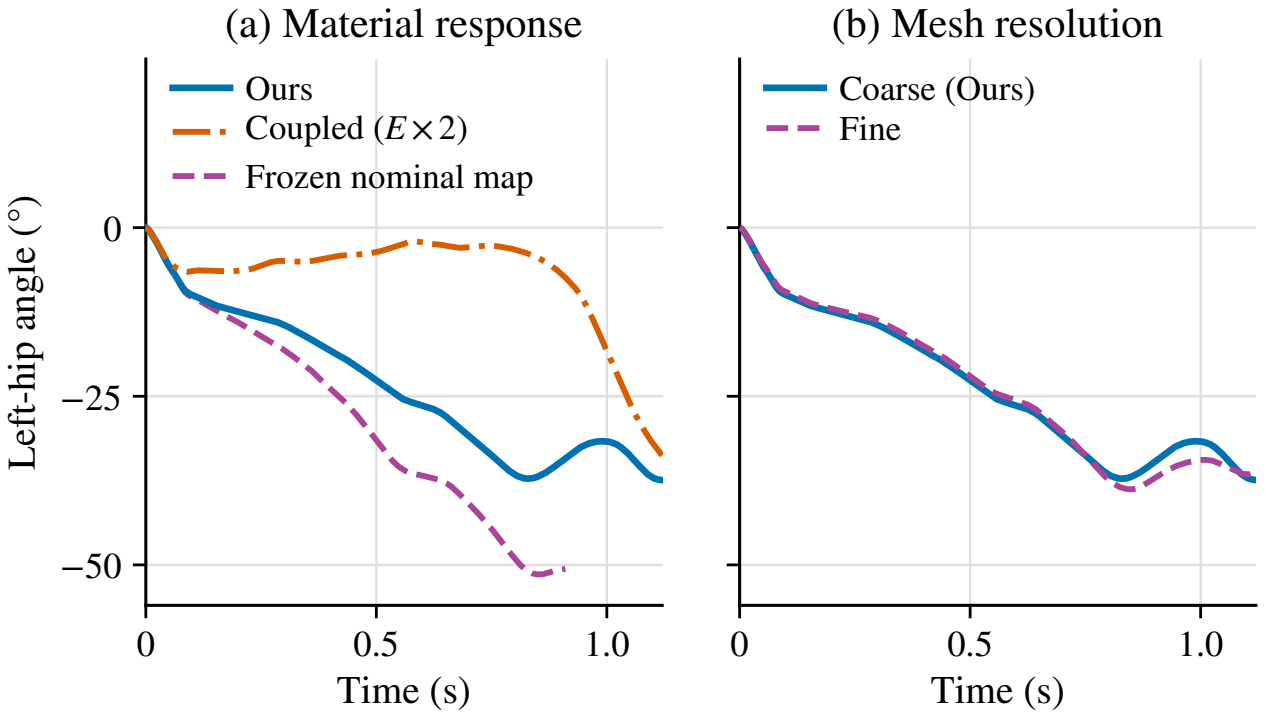}
\caption{\textbf{Simulation with identical activation sequences.} Left-hip angles from matched initial skeletal states: (a) \ours{} at nominal stiffness, coupled simulation with doubled Young's modulus, and the frozen nominal map; (b) coarse 9-tet versus fine 230-tet meshes. The frozen-map rollout stops at 0.91\,s when its inputs exceed the calibrated range.}
\label{fig:mechanical_response}
\end{figure}

In Figure~\ref{fig:mechanical_response}, we simulate the same recorded aiming activation sequence from identical initial skeletal states. Panel (a) compares \ours{} at nominal stiffness, coupled simulation with doubled Young's modulus ($E\times2$), and the frozen nominal map. Active strength and tendon parameters are unchanged. The frozen-map rollout stops at 0.91\,s when its inputs exceed the calibrated range. Panel (b) compares coarse 9-tet and fine 230-tet meshes under the same activations. Table~\ref{tab:feedback_ablation} reports fixed-policy tracking after controlled belly-shape or material changes, with strength scales, tendon parameters, skeleton, routing, contacts, observations, and goals held fixed.

\begin{table}[t]
\centering\small
\setlength{\tabcolsep}{3pt}
\renewcommand{\arraystretch}{0.95}
\caption{\textbf{Tissue-response feedback under material change.}
A fixed policy is evaluated after doubling Young's modulus.
Paired values follow the Ours / Frozen order.
$\Delta_Q$ measures frozen-load discrepancy relative to Ours.}
\label{tab:feedback_ablation}
\begin{tabular}{@{}llccc@{}}
\toprule
Change & Method & $\Delta_Q$ (\%)
& Local error (cm) & Success (\%) \\
\midrule
$E\times2$ & Ours / Frozen & 11.48
& 9.71 / 8.89 & 68.0 / 76.1 \\
\bottomrule
\end{tabular}
\end{table}

\subsection{Independent actuation-model ablation}

Table~\ref{tab:coupling} compares Hill-type and coupled volumetric actuation. Each controller and world model is trained independently using the same model-based learning procedure and network architectures. We match the skeleton, actuator correspondence, nominal strength, motion references, rewards, initial-state distribution, control rate, policy observations, and physical-transition and optimization-update budgets. Wall-clock cost is reported separately. Backend-specific internal states, including volumetric nodal coordinates, are excluded from policy and world-model inputs.

\begin{table}[t]
\centering\small
\caption{\textbf{Actuation-model ablation.} Both backends are trained independently with matched learning/evaluation conditions. Local error is in cm, success in percent, and throughput in aggregate physical control transitions/s.}
\label{tab:coupling}
\begin{tabular}{@{}lccc@{}}
\toprule
Actuation & Local error & Success & Samples/s\\
\midrule
Hill line & 4.92 & 97.9\% & 9030\\
Volumetric & 5.52 & 97.9\% & 815\\
\bottomrule
\end{tabular}
\end{table}

We evaluate tracking on 47 clips from the LAFAN1 motion dataset, initializing all compared methods from the same reference start states and using deterministic policy rollouts. Metrics include root, local-body, and endpoint errors, tracking success rate, normalized return, activation magnitude and variation, and training cost. An episode terminates early if the absolute root-height error exceeds $0.25\,\mathrm{m}$, the root-tilt discrepancy $\lvert R_{zz}-R^{\mathrm{ref}}_{zz}\rvert$ exceeds $0.8$, or horizontal root-position error exceeds $10\,\mathrm{m}$. A rollout is considered successful if it reaches $10\,\mathrm{s}$ or the end of the reference clip, whichever occurs first, without triggering a failure termination.

\subsection{World-model learning efficiency}
\label{sec:world_model_learning_efficiency}

Table~\ref{tab:learning_ablation} compares our world-model training at critic horizons of 16, 32, and 48 steps with three baselines: PPO~\cite{ppo}; MS-Emulator~\cite{wei2026msemulator}, model-free adversarial imitation with a value-guided flow policy; and PPO-WM, which keeps our world model, imagined rollouts, rewards, and critic but updates the policy with PPO. All variants use the same motion split, reward, common observations, policy/critic family, and physical-transition budget.

\begin{figure}[t]
\centering
\input{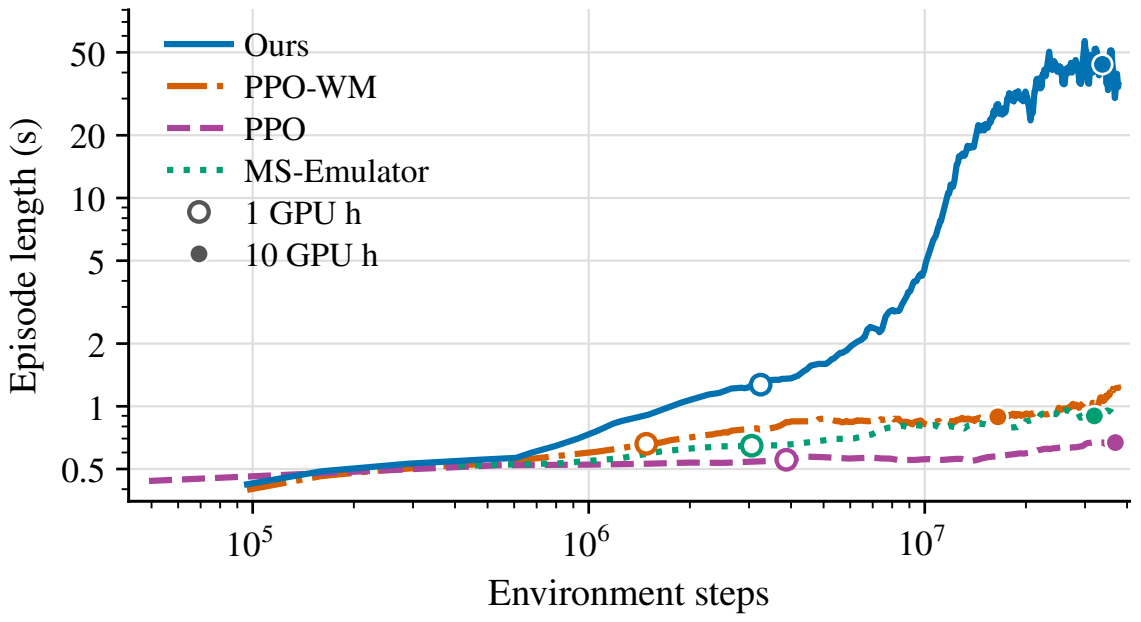}
\caption{\textbf{Learning efficiency in the volumetric simulator.} Mean episode length is reported in seconds. Transitions (millions) and training hours to reach a mean episode length of $\tau=10\,\mathrm{s}$ are reported separately. Open and filled dots mark where each method had used 1 and 10 GPU hours.}
\label{fig:learning_efficiency}
\end{figure}

At the default 32-step horizon, mean training episode length reaches 38.2 s with a normalized per-step reward of 0.709.
Figure~\ref{fig:learning_efficiency} plots mean episode length against physical transitions and training time, including collection and all learning updates. We report transitions and hours to reach a mean episode length of 10 s; runs below this threshold are censored at their budget.
On a shared fixed replay set with identical evaluation normalization, four-step world-model state prediction MSE is 5.283, 3.798, and 3.688 for horizons of 16, 32, and 48 steps, respectively.

\begin{table}[t]
\centering\small
\setlength{\tabcolsep}{4pt}
\caption{\textbf{Learning efficiency in the volumetric simulator.} Mean episode length is reported in control steps. Transitions (millions) and training hours to reach a mean episode length of $\tau=10\,\mathrm{s}$ are reported separately. World-model labels indicate critic horizons in control steps.}
\label{tab:learning_ablation}
\begin{tabular}{@{}lccc@{}}
\toprule
Variant & Ep. length & Trans.@$\tau$ & Hours@$\tau$\\
\midrule
PPO & 31.6    & $>$188.7 & $>$50.2\\
Critic horizon: 16 & 503.4 & 11.616 & 3.64\\
Critic horizon: 32 & 506.5 & 9.632 & 3.00\\
Critic horizon: 48 & 506.5 & 11.487 & 3.56\\
\bottomrule
\end{tabular}
\end{table}

\subsection{Muscle recruitment}
\label{sec:recruitment}

\begin{figure}[t]
  \centering
  \includegraphics[width=\linewidth]{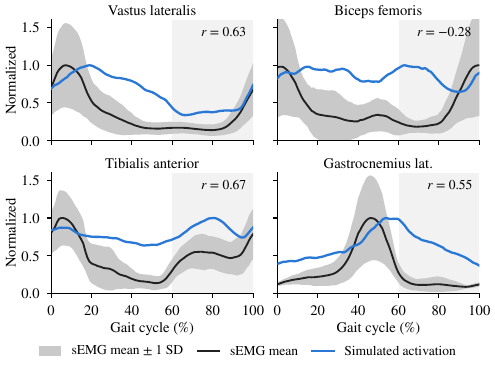}
  \caption{Simulated activation (blue, mean of 97 strides) versus sEMG
  (black: mean; gray: $\pm$1 SD; 13 adults at 4\,km/h~\cite{moreira2021lower})
  over the gait cycle, right leg, peak-normalized; 0\% is heel strike, shading
  marks typical swing (60--100\%). $r$: Pearson correlation of the mean curves.}
  \label{fig:emg}
\end{figure}

We compare activations during LaFAN1 walking with surface
EMG~\cite{moreira2021lower}. Each of the 12 walking clips is played from its
first frame for up to 180\,s; when the character falls, it restarts from the
reference state at that frame, and the first second after each restart is
discarded, which leaves 97 right-leg strides. For each electrode we take the
force-weighted mean activation of the synergist group under it
(Figure~\ref{fig:emg}). Without EMG supervision, the coarse timing of
recruitment emerges: the quadriceps is most active in early stance
($r = 0.63$), the triceps surae in push-off ($r = 0.55$, 11\% of the cycle
late), and the tibialis anterior follows the stance-to-swing pattern
($r = 0.67$), though without its burst after heel strike. The modulation does
not: between bursts the simulated groups stay at 34--65\% of their peak,
against 9--19\% in sEMG, so only 26\% of the cycle lies within one standard
deviation of the human mean, and the hamstrings remain active through stance.
For the single muscle under each electrode, the mean correlation is $r = 0.48$.
Kinematic tracking alone does not determine how work is split among synergists.

\subsection{Limitations}

The coarse belly approximates anatomy. Direct endpoints are spring approximations, while segment/cylinder wrapping permits finite penetration and omits explicit fascia and general tissue--tissue contact. Quasi-static tissue omits belly inertia, oscillations, and impulse transmission; the fiber response omits force--velocity effects. Finite equilibrium residuals and held-force stepping can affect work accuracy. Motion tracking alone does not resolve muscle-recruitment ambiguity or establish physiological validity~\cite{hicks}.

\section{Conclusion}
\label{sec:conclusion}

We present \ours{}, a model-based reinforcement learning framework for full-body motion control with volumetric muscles and two-way muscle--skeleton coupling. Bone pose changes attachment and wrapping energy, while tissue equilibrium supplies skeletal reactions. Simplified tetrahedral muscle meshes make volumetric simulation efficient enough for learning. Our method achieves accurate full-body motion tracking on LaFAN1, custom fitness sequences, and motions reconstructed from videos.

\label{main-text-end}
\label{backmatter-start}
{\small
\nocite{gae}
\bibliographystyle{ieeenat_fullname}
\bibliography{main}}

\begin{bibunit}[ieeenat_fullname]
  \backrefsetup{disable}
  \clearpage
\appendix

\setcounter{figure}{0}
\setcounter{table}{0}
\setcounter{equation}{0}
\renewcommand{\thefigure}{S\arabic{figure}}
\renewcommand{\thetable}{S\arabic{table}}
\renewcommand{\theequation}{S\arabic{equation}}
\begin{center}
  {\large\bfseries \papertitle}\\[4pt]
  {\large Supplementary Material}
\end{center}

\section{Muscle Model Construction and Mechanics}

\label{app:force}

We use anatomical volumes to size the muscle bellies. Bone-mounted attachments, routed tendons, and wrapping cylinders transmit their forces to the skeleton.

\subsection{Volumetric Asset Construction}
\label{app:assets}

\subsubsection{MASS284-to-RLMuscleV2 Asset Construction}

\begin{figure}[t]
\centering
\includegraphics[width=\linewidth]{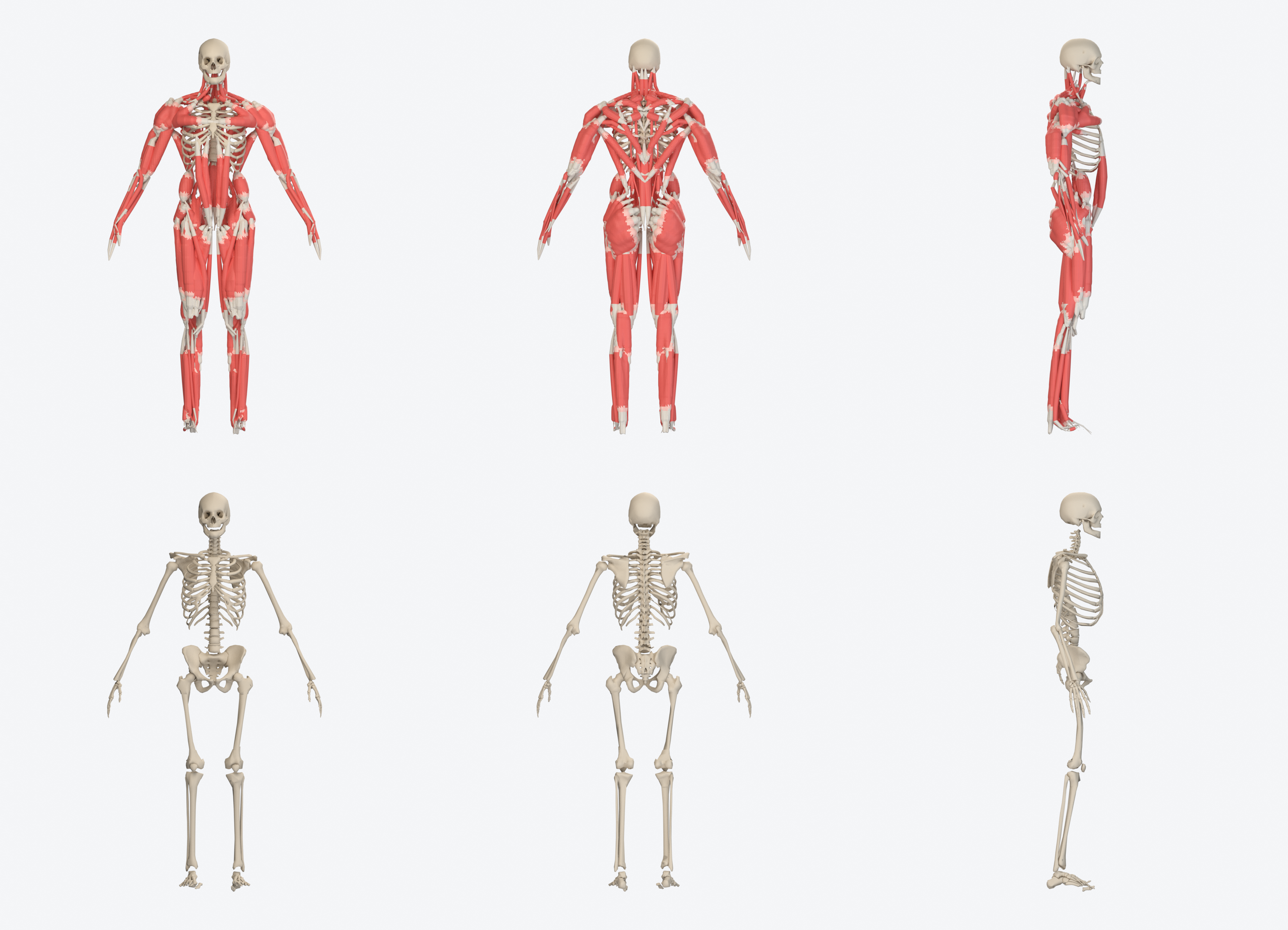}
\caption{Anatomical calibration of the MASS284 muscle geometry.}
\label{fig:anatomy}
\end{figure}

We start with the 284 muscle paths in MASS284\cite{mass} and keep the original skeleton, bone waypoints, and actuator order. We turn each path into a tetrahedral tube and use the RLMuscleV2\cite{rlmuscle} anatomical meshes to set its size (Fig.~\ref{fig:anatomy}). VIPER~\suppcite{angles2019viper} also samples cross-sections along a muscle, connecting points across slices to form a bundle of elastic rods.

\begin{figure}[!hb]
\centering
\includegraphics[width=\linewidth]{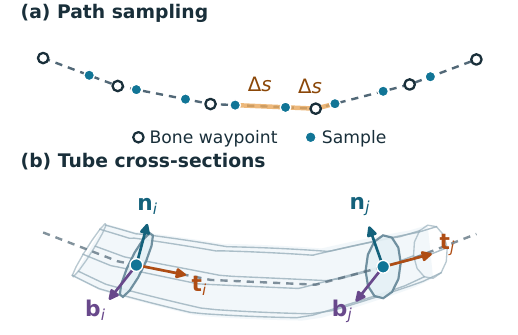}
\caption{\textbf{Sampling the MASS284 path and orienting tube cross-sections.}
(a) The dashed line joins the original bone waypoints (open circles). We place samples (filled circles) every $\Delta s$ of distance measured along the polyline.
(b) The blue cross-sections are drawn at samples $i$ and $j$. At each sample, $\mathbf t$ points along the path, and $\mathbf n$ and $\mathbf b$ span the cross-section.}
\label{fig:supp-tube-frame}
\end{figure}

\paragraph{Building a tube along the path.}
We sample each path at equal distances and place a cross-section at each sample (Fig.~\ref{fig:supp-tube-frame}). At sample $i$, $\mathbf t_i$ points along the path, and $\mathbf n_i$ and $\mathbf b_i$ are perpendicular directions in the cross-section plane. As the path turns, we apply the smallest rotation $R_i$ that maps $\mathbf t_i$ to $\mathbf t_{i+1}$. This gives $\mathbf n_{i+1}=R_i\mathbf n_i$ and $\mathbf b_{i+1}=\mathbf t_{i+1}\times\mathbf n_{i+1}$, keeping successive cross-sections aligned through bends.

The cross-sections are narrow at the ends and widest in the middle. Let $s$ be the distance along the path from its starting point and $L$ the total path length. The radius at $s$ is
\begin{equation}
 r(s)=\tfrac12\rho L\left[0.25+0.75\sin^{0.8}(\pi s/L)\right].
\end{equation}
Here $\rho=2r(L/2)/L$ is the midpoint diameter divided by path length. Each calibration tube has ten 12-sided cross-sections and two end tips, giving 122 vertices and 290 tetrahedra per unit.

A separate high-resolution fusiform model with explicit tendon material assigns active contraction to the central $64\%$ of the rest path, with a passive tendon spanning 18\% of the path at each end.

\paragraph{Setting the muscle volumes.}
We assign each MASS284 path to an anatomical muscle group in RLMuscleV2's \texttt{humanV2} asset~\cite{rlmuscle} by voting over the nearest mesh vertices along the path. We measure the group's reference volume $V_G$ by summing its tetrahedra in the rest pose. When several MASS284 units map to the same group, we split $V_G$ in proportion to their nominal Hill strengths. Here $m$ and $j$ index units assigned to group $G$, and $f_{0,m}$ is the nominal strength of unit $m$. The initial share is $V_m^{\mathrm{raw}}$. We apply a common volume scale and copy the targets of paired left units to the right to obtain the final target $V_m$ for each unit. The volume share and effective cross-sectional area are
\begin{equation}
\begin{aligned}
 V_m^{\mathrm{raw}}&=V_G\frac{f_{0,m}}{\sum_{j\in G}f_{0,j}},\\
 A_{\mathrm{eff},m}&=\frac{V_m}{L_{f,m}}.
\end{aligned}
\end{equation}
We estimate the fiber span $L_{f,m}$ from the matched anatomical group by projecting its vertices onto the mean fiber direction and taking the span between the 5th and 95th percentiles, with a $2\,\mathrm{cm}$ minimum.

We use $A_{\mathrm{eff},m}$ only to initialize the radius of the fusiform calibration mesh. A regular 12-sided cross-section with circumradius $r$ has area $3r^2$. Setting the midpoint cross-sectional area to $A_{\mathrm{eff},m}$ and using $r(L/2)=\rho L/2$ gives the initial radius scale
\[
 \rho_0=\frac{2}{L}\sqrt{\frac{A_{\mathrm{eff},m}}{3}}.
\]
We clamp $\rho_0$ to $[0.02,0.35]$ before adjusting $\rho$ to match the target rest volume $V_m$.

For the 9-tet mesh, the belly width is $w_m=\sqrt{3V_m/(2L_m)}$, where $L_m$ is the straight-line distance between the two belly tips. In the elongated 9-tet variant, we use $64\%$ of the total rest polyline length as the default target for $L_m$ when selecting the tip locations.

The 284 calibration tubes have a combined rest volume of $30.24\,\mathrm{L}$, summed over their muscle tetrahedra. The anatomical volume calibration excludes the organ group. For simulation, we replace each tube with a belly made of two quadrilateral cross-sections and two tips, with fibers along the belly. We scale the coarse belly to the corresponding tube's volume, so the two mesh sets have the same combined rest volume. The outer parts of the original paths remain as tendon waypoints. Selected shoulder and hip bellies are curved or flattened to fit their routes and the nearby bones. We rescale their cross sections to match the target volume and check that all rest tetrahedra have positive signed volume. Table~\ref{tab:meshes} gives the full-body mesh counts. Figure~\ref{fig:supp-muscle-mesh-variants} compares the three discretizations on one muscle.

\begin{table}[t]
\centering\small
\caption{Full-body MASS284 meshes for geometry calibration and simulation.}
\label{tab:meshes}
\begin{tabular}{@{}lrrr@{}}
\toprule
Mesh & Units & Vertices & Tetrahedra\\
\midrule
290-tet tube & 284 & 34,648 & 82,360\\
9-tet diamond & 284 & 2,840 & 2,556\\
\bottomrule
\end{tabular}
\end{table}

Each endpoint uses its configured routed tendon or direct attachment. For a tetrahedron $e$ in muscle unit $m$, the material and active weights are $w_e=V_e^0Y_m/(1+\nu)$ and $c_e=V_e^0S_m$. Here $V_e^0$ is the rest volume, $Y_m$ is the fitted Young's modulus, and $S_m$ is the active-fiber coefficient. The relative weight of the volume-preserving term is $\nu=0.3$. We use the MASS284 Hill-type line-muscle model as the calibration reference. At sampled skeletal poses and activation levels, we evaluate each line muscle's force at zero contraction velocity and compute its joint torques from its moment arms. We solve the volumetric muscles at the same poses and activations and fit $Y_m$, $S_m$, attachment stiffness, and routed-tendon slack lengths to match the reference forces and torques. We select the fitted parameters on held-out poses.

\subsubsection{Surface Geometry}

For visualization, we merge the bellies of units that belong to the same anatomical muscle into a continuous surface. Let $f_m(x)$ be the implicit surface field of tube $m$ at position $x$. We blend these fields with a smooth minimum using $k=100\,\mathrm{m}^{-1}$.

\begin{equation}
 f_{\rm union}(x)=-\frac1k\ln\sum_m\exp[-k f_m(x)].
\end{equation}

The blend weight is zero over the outer $20\%$ of the path at each end and increases smoothly over the next $8\%$ of the path length, keeping the tendon ends separate. This step affects only the rendered surface. The simulation uses the original tetrahedral mesh and contact constraints.

\subsection{Tendon Routing and Direct Attachments}

For one routed endpoint, write its chain as $p_0,\ldots,p_n$, beginning at the muscle tip and continuing through bone-fixed waypoints. Segment directions and tension are
\begin{equation}
 d_j=\frac{p_{j+1}-p_j}{\|p_{j+1}-p_j\|},\qquad
 T=k_s[L-L_0]_+.
\end{equation}
Here $j$ indexes chain segments, $d_j$ is their unit direction, $L$ is total routed length, $L_0$ is slack length, and $k_s$ is tendon stiffness. The positive-part operator gives zero tension for a slack route. Differentiating its energy gives
\begin{equation}
 f_0=Td_0,\quad f_j=T(d_j-d_{j-1}),\quad f_n=-Td_{n-1}.
\end{equation}
Here $f_0$ acts on the tissue tip, $f_j$ acts on interior bone waypoint $j$ for $1\leq j<n$, and $f_n$ acts on the terminal bone attachment. Each interior waypoint receives the vector sum of the pulls from its two adjacent segments. A direct attachment is the single-segment case ($n=1$) with zero slack length ($L_0=0$).

\subsection{Cylinder Wrapping and Contact Reactions}

\begin{figure}[t]
\centering
\includegraphics[width=\linewidth]{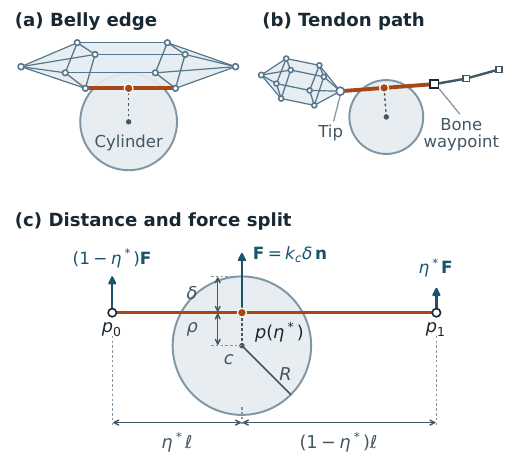}
\caption{\textbf{Cylinder wrapping and contact reactions.}
Views along the cylinder axis. The highlighted segment is a muscle-mesh
edge in (a) and the tip-to-waypoint span of a routed tendon in (b).
In (c), the closest point $p(\eta^*)$ sets the penetration $\delta$ and
the force weights on $p_0$ and $p_1$. The segment length is
$\ell=\|p_1-p_0\|$.}
\label{fig:supp-cylinder-wrapping}
\end{figure}

Each muscle with wrapping checks 20 outer belly edges and up to two segments connecting a muscle tip to its first bone waypoint (Fig.~\ref{fig:supp-cylinder-wrapping}(a,b)). For one segment and cylinder,
\begin{equation}
\begin{aligned}
 p(\eta)&=(1-\eta)p_0+\eta p_1,\qquad 0\leq\eta\leq1,\\
 \rho&=\min_{\eta\in[0,1]}
 \|(I-d_cd_c^\top)(p(\eta)-c)\|,\\
 \delta&=[R-\rho]_+.
\end{aligned}
\end{equation}
Here $p_0,p_1$ are segment endpoints, $\eta$ is the interpolation coordinate, $d_c$ is the unit cylinder axis, $c$ is a point on that axis, and $I$ is the identity matrix. The projector $I-d_cd_c^\top$ removes axial displacement, $\rho$ is the closest radial distance, $R$ is effective cylinder radius including clearance, and $\delta$ is penetration depth.

We split the contact force at $p(\eta^*)$ between the two endpoints (Fig.~\ref{fig:supp-cylinder-wrapping}(c)) using
\begin{equation}
 f_0=(1-\eta^*)k_c\delta n,\qquad
 f_1=\eta^* k_c\delta n.
\end{equation}
Here $\eta^*$ is the minimizing segment coordinate, $n$ is the outward radial unit normal, $k_c$ is penalty stiffness, and $f_0,f_1$ are endpoint contact forces. If an endpoint belongs to a bone, its contribution is applied to that bone. The cylinder's owning bone receives the opposite force and torque. Degenerate radial normals require a configured fallback direction. This soft segment test permits penetration and does not guarantee collision-free triangle interiors or fascia interaction.

\paragraph{Model assignments.}
Table~\ref{tab:supp-wrapping-assignment} counts individual ends by connection type and assigns each muscle unit to one endpoint category. A unit counts as direct if both ends attach directly to bone and as routed if at least one end uses a routed tendon. The cylinder wrapping row counts wrapped units independently of their endpoint connections.

\begin{table}[t]
\centering\small
\caption{Endpoint and cylinder-contact assignments in the wrapping-enabled model asset used for training. Fig.~\ref{fig:muscle-wrapping} in the main text illustrates the three configurations. Each unit is one muscle belly with an origin and an insertion. The tibialis anterior has a direct origin and a routed insertion. Deltoid and gluteus minimus each have three wrapped units per side.}
\label{tab:supp-wrapping-assignment}
\setlength{\tabcolsep}{3pt}
\begin{tabular*}{\linewidth}{@{\extracolsep{\fill}}lrrl@{}}
\toprule
Assignment & Ends & Units & Representative muscles \\
\midrule
Direct spring & 178 & 14 & Piriformis \\
Routed tendon & 390 & 270 & Biceps brachii, soleus \\
Cylinder wrap & -- & 12 & Deltoid, gluteus minimus \\
\bottomrule
\end{tabular*}
\end{table}

\section{Simulation Engine and GPU Execution}

At each tissue update, we batch the coarse tetrahedral muscle meshes on the GPU and solve for quasi-static equilibrium at the current skeletal pose. We use Projective Dynamics to compute muscle deformation and run an additional contact solve for muscles that wrap around cylinders. The rigid-body integrator then advances the skeleton over several substeps using the resulting muscle loads. The muscle solve omits tissue inertia and gravity, while rigid-body integration accounts for skeletal inertia and gravity.

\subsection{Runtime Muscle Meshes and GPU Batching}

The simulator solves for quasi-static muscle equilibrium at the current skeletal pose and holds the resulting loads across rigid-body substeps. The muscle solve omits tissue inertia and gravity, while the skeletal dynamics retain both.

The simulator solves for the shape of all muscles at every tissue update during rollout collection. We use a coarse mesh to reduce the number of element projections and vertex displacements per iteration. Each belly has ten vertices and nine tetrahedra, arranged as two quadrilateral cross sections and two tips. This reduces the tetrahedron count per environment from 82,360 in the calibration mesh to 2,556 at runtime (Table~\ref{tab:meshes}). With the same block size for all muscles, the GPU can batch the solves across muscles and environments. The coarse mesh retains the belly span, target volume, reference fiber directions, and bone-local tendon waypoints.

\subsection{Projective Dynamics and Wrapping Solves}

Algorithm~\ref{alg:volumetric-simulator} describes one simulator control step. We read the skeletal pose and velocity and hold the policy's muscle activations fixed for that step. At each tissue update, we refresh the attachments, tendon waypoints, and wrapping cylinders from the current pose. We first run the muscle solve for all units in parallel, initializing each unit $m$ with $x_m^{-}$, the most recently accepted 3D positions of its ten mesh vertices. We then run a wrapping solve for the units with configured cylinders $C_m$.

With the skeleton and activation fixed, the equilibrium solve moves the muscle vertices until the net force on each vertex is small. We use Projective Dynamics (PD)~\cite{projectivedynamics} because each iteration can update the nonlinear material response locally while reusing the base Cholesky factorization for the vertex solve. For each tetrahedron, we compute its deformation gradient and volume ratio, then find a local rotation and a volume-preserving projection. We combine the resulting material forces with the forces from the active fiber energy (Eq.~\ref{eq:fiberprofile}), tendons, attachments, and the volume barrier to form the nodal force residual. In the muscle solve, we use a $10\times10$ system with the same factorization for all three coordinate components, using rank-one endpoint corrections for routed tendons.

The wrapping solve adds cylinder-contact energy. After a reset, the muscle solve provides the initial candidate. On subsequent updates, we start from the last accepted wrapping shape moved with the tendon leads. A contact-aware initialization can adjust either candidate before the wrapping iterations. We use the accepted wrapping shape to compute the loads of those muscles. Backtracking uses up to 40 trials per iteration, halving the step size after each rejection. A trial must pass the Armijo test with coefficient $10^{-4}$ and satisfy $J_e=\det F_e>J_{\min}$ in every tetrahedron. The wrapping solve reuses determinant polynomials to check the volume bound along the search direction. Iteration stops at the residual tolerance or iteration cap.

\subsection{Multirate Muscle--Rigid Body Coupling}
The policy supplies activation at $50\,\mathrm{Hz}$. At each scheduled tissue update ($100\,\mathrm{Hz}$), the solver refreshes tendon waypoints, attachment targets, and wrapping geometry from the current skeletal pose, solves for quasi-static muscle equilibrium, and maps the resulting reactions to a generalized muscle load. The fixed-hold configuration reuses this load for ten rigid-body substeps at $1000\,\mathrm{Hz}$ in the Warp implementation of MuJoCo~\suppcite{todorov2012mujoco}, so the tissue solve and the force mapping run only at tissue updates. The backend also supports an earlier refresh when pose changes exceed a configured threshold. With fixed hold enabled, the LaFAN1 tracking run achieved approximately $800$ physical control transitions per second across 2,048 parallel environments on one NVIDIA GeForce RTX 4090. This end-to-end rate includes the time spent updating the world model and actor--critic, which reduces the number of simulation transitions collected per second. The run used 284 muscle units and 12 configured cylinder-wrapping pairs per environment. The nodal residual target was $1\,\mathrm{N}$, with iteration caps of 600 for the muscle solve and 50 for the wrapping solve. The wrapping configuration also accepted finite iterates above the volume threshold $J_{\min}=0.1$ when the residual target was unmet. In this window, $95.1\%$ of wrapping blocks met the residual target.

\section{Reference Motion Processing}

Motion-capture clips and video-derived poses provide the tracking references. We convert both sources into joint trajectories for the tracking tasks.

\subsection{LaFAN1 References and Motion Conversion}

We train on motion clips from LaFAN1~\cite{lafan}. For evaluation, we retain 47 clips from this training set after excluding \texttt{ground}, \texttt{fallAndGetUp}, and \texttt{pushAndFall} clips.

Before resampling the reference motions, we unwrap each joint-angle channel by taking the shortest signed angular increment between consecutive frames.
\begin{equation}
 q^u_{t+1}=q^u_t+\operatorname{wrap}_{(-\pi,\pi]}(q_{t+1}-q_t).
\end{equation}
We linearly interpolate the unwrapped angles at the control rate and compute joint velocities from that continuous sequence. This avoids spurious velocity spikes when an angle crosses the $\pm\pi$ boundary.

\subsection{AMASS and Motion-X References}

We use flat-ground motion clips from the CMU, SFU, and TotalCapture subsets of AMASS\suppcite{mahmood2019amass}. We train on 1,758 clips and hold out 159 clips for validation and 283 for testing.

We combine 1,874 indexed clips with 326 additional clips from local recordings. For the indexed clips, we use captions from Motion-X\suppcite{motion-x} and HumanML3D\suppcite{humanml3d}. For the additional clips, we use the original trial descriptions and take names. We retain clips that pass pose, ground, and kinematic checks. The full set contains 2,200 clips and 555,108 frames at $30\,\mathrm{Hz}$, totaling $5.14$ hours.

\subsection{Video Pose Estimation and Retargeting}

We extract 2D image landmarks and pelvis-relative 3D body landmarks using the body-pose estimator from BlazePose GHUM Holistic~\cite{mediapipepose}, implemented as MediaPipe Pose Full. We adapt the landmarks to the bone lengths and recover joint angles with joint-constrained inverse kinematics, followed by temporal smoothing. For walking clips, we estimate root translation from stance-foot motion. We resample the trajectories at $50\,\mathrm{Hz}$ and convert them to the policy skeleton, preserving root poses and checking joint limits and forward-kinematic consistency.

For strength training motions, direct frame-by-frame pose estimation from video can produce noisy trajectories. We therefore select representative poses from demonstration videos, adapt these key poses to the target skeleton, and smoothly interpolate between them to construct continuous motion references for exercises such as dumbbell lateral raises and deadlifts.

We also retarget Reallusion sword-and-shield clips distributed with ASE~\suppcite{peng2022ase} and backhand motion from Tennis-MoCap~\suppcite{pulgarin2017tennis} to the target skeleton. The tennis reference is resampled to $50\,\mathrm{Hz}$ and fitted with joint limits and two-hand racket-grip constraints.

\section{Policy and World-Model Training}
\label{app:tracking}

We fit the world model to transitions collected in the coupled simulator. The policy and critic learn from its imagined rollouts using rewards defined by the reference motion. The same training pipeline also supports task-specific goals and rewards.

\subsection{Training Pipeline}

MuscleVAE~\cite{musclevae} supplies the learned-dynamics control precedent, and we use a GPU-accelerated implementation of SuperMimic~\cite{supermimic} for shared replay, actor, world-model, and critic updates. We use this learning procedure for volumetric muscle control without the generative VAE component.

Algorithm~\ref{alg:volumetric-training} summarizes the training loop. In each iteration, we run the policy in the coupled simulator to collect transitions, fit the model to these transitions, and then update the actor and critic on trajectories predicted by the fitted model. The stepping rates are given in Section~\ref{sec:multi_rate}, and the network sizes and training settings are listed in Table~\ref{tab:supp-training-hyperparameters}.

\begin{table}[!hb]
\centering\small
\caption{\textbf{Volumetric tracking hyperparameters.} Settings for the volumetric policy. Network widths list hidden layers.}
\label{tab:supp-training-hyperparameters}
\begin{tabular}{@{}lr@{}}
\toprule
Setting & Value \\
\midrule
Policy encoder widths & $2048, 1024, 512, 256$ \\
Policy decoder widths & $2048, 1024, 512, 256$ \\
Goal latent / action size & $32 / 284$ \\
World-model widths & $1024, 1024, 1024$ \\
Critic widths & $256, 128$ \\
\addlinespace[3pt]
Parallel environments & 2,048 \\
Replay per environment & 500 transitions \\
Total replay capacity & 1,024,000 \\
Policy / world-model batch & 2,048 each \\
World-model training window & 4 control steps \\
Policy / critic horizon & 16 / 32 control steps \\
World-model batches / iteration & 5 \\
World-model-only warm-up & 200 iterations \\
\addlinespace[3pt]
Base learning rate & $4\times10^{-4}$ \\
Policy / model optimizer & Muon + Adam \\
Critic optimizer & Adam \\
Discount $\gamma$ / return $\lambda$ & $0.98 / 0.90$ \\
Policy EMA decay & 0.995 \\
Exploration at $a=0.5$ & Beta, $\sigma=0.0707$ \\
\bottomrule
\end{tabular}
\end{table}

\subsection{World-Model Fitting and Policy Optimization}

We fit the world model by unrolling it over recorded activation sequences from replay and comparing its predictions with the simulator states, using a four-step window. Given the current state and the current and next muscle activations, the model predicts the next root position and orientation, root linear and angular velocities, joint angles and velocities, and normal ground-contact forces on the foot bodies. In the direct-activation setting, the next activations are the 284 excitations clipped to $[0,1]$. Muscle deformation and forces are not predicted by the model. They are computed only by the physical solver during data collection.

Each replay transition also stores two flags. The \texttt{done} flag stops the unroll at the end of an episode, and the \texttt{valid} flag excludes rejected muscle solves and non-finite transitions from world-model fitting.

The policy takes the current physical observation and the previous excitation as input, without stacking past frames. Its goal consists of the motion phase, the current reference frame, and four future reference frames spaced $0.2\,\mathrm{s}$ apart.

To update the actor, we freeze the world model and backpropagate the imagined return in Eq.~\eqref{eq:training} through its predicted states and its action inputs, so the gradients pass through the model Jacobians instead of stopping at each predicted state. Each imagined reward is weighted by the product $\prod_{j<k}c_j$ of the preceding soft continuation probabilities $c_j=1-d_j$, where $d_j$ combines the soft termination terms of Appendix~\ref{sec:tracking_reward_terms}. At every imagined step, we also clip the gradient carried to the next step to a norm of 100 per trajectory, so that a few trajectories cannot dominate an update. Appendix~\ref{sec:supp-horizon} discusses how the horizon affects these gradients.

The critic is trained on the same imagined rollouts. Its targets are finite-horizon $\lambda$-returns~\cite{gae}:
\begin{equation}
\begin{split}
 G_H^\lambda&=V_\psi(\hat s_H,g_H),\\
 G_k^\lambda&=r_k+\gamma c_k\left[(1-\lambda)
 V_\psi(\hat s_{k+1},g_{k+1})+\lambda G_{k+1}^\lambda\right],
\end{split}
\end{equation}
computed backward from $k=H-1$ to $0$. Bootstrapped values in both the actor objective and these targets are clamped to the largest attainable return, $r_{\max}/(1-\gamma)$, where $r_{\max}$ is the maximum per-step reward. Table~\ref{tab:supp-training-hyperparameters} lists the horizon, $\gamma$, and $\lambda$.

\subsection{Tracking Objective}
\label{sec:tracking_reward_terms}

Local articulation is compared in heading frames, while global root trajectory remains explicit. Let $p$ and $H$ be simulated root position and yaw rotation, and let barred symbols denote reference quantities. For body $i$, define
\begin{equation}
\begin{split}
 z_i^p&=H^\top(p_i-p),\quad
 \bar z_i^p=\bar H^\top(\bar p_i-\bar p),\\
 e_i^v&=H^\top v_i-\bar H^\top\bar v_i,\\
 e_i^\omega&=H^\top\omega_i-\bar H^\top\bar\omega_i.
\end{split}
\end{equation}
Orientations are expressed in the same heading-aligned frame. Each velocity must be rotated by its own heading: if we rotated both the reference and simulated velocities by the simulated heading alone, any yaw offset between them would persist as a world-frame mismatch. The rotated velocities are instantaneous vectors expressed in this frame, not time derivatives of root-relative positions. As a result, rigidly rotating the entire reference about the vertical axis leaves all local terms unchanged. Only the root-heading term changes, and it captures the relative yaw between the reference and the simulated character.

The objective uses exponential body-error kernels and separate root terms. With horizontal root distance $d$, wrapped yaw error $\delta\psi$, and endpoint set $\mathcal E$,
\begin{equation}
\begin{split}
 K_\sigma(d)&=\exp(-d^2/\sigma^2),\\
 r_{\rm root}&=\max(0,1-d/D),\\
 r_{\rm yaw}&=\exp[-\rho_\delta(\delta\psi/\sigma_\psi)],\\
 r_{\rm tip}&=|\mathcal E|^{-1}\sum_{i\in\mathcal E}
 \exp(-\kappa\|z_i^p-\bar z_i^p\|^2).
\end{split}
\end{equation}
The root tent has constant radial slope inside its support and zero slope outside. For the yaw kernel, let $z=\delta\psi/\sigma_\psi$ and $c=2$. Then $\rho_\delta(z)=z^2$ for $|z|\leq c$ and $\rho_\delta(z)=c^2+2c(|z|-c)$ otherwise. The endpoint score averages two fingertip scores and one head-top score.

\begin{table}[t]
\centering\small
\caption{Shared tracking terms in the endpoint configuration. The continuous weighted sum is multiplied by the control interval $\Delta t$.}
\label{tab:supp-tracking-rewards}
\begin{tabular}{@{}lcl@{}}
\toprule
Component & Weight & Scale / form \\
\midrule
Body position & 1.0 & $\sigma=0.3\,\mathrm{m}$ \\
Body orientation & 1.0 & $\sigma=0.4\,\mathrm{rad}$ \\
Local linear velocity & 0.25 & $\sigma=1.0\,\mathrm{m/s}$ \\
Local angular velocity & 0.25 & $\sigma=3.14\,\mathrm{rad/s}$ \\
Root position & 0.5 & Tent radius $4\,\mathrm{m}$ \\
Root yaw & 1.0 & $\sigma_\psi=0.25\,\mathrm{rad}$ \\
Hand/head endpoints & 1.5 & $\kappa=1000\,\mathrm{m}^{-2}$ \\
Action rate & 0.03 & $-\|a_t-a_{t-1}\|_2^2$ \\
\bottomrule
\end{tabular}
\end{table}

Optional arm and trunk orientation terms have zero weight in the archived endpoint configuration. The positive tracking weights sum to 5.5, giving a maximum step score of 0.11 at 50\,Hz. The configured hard termination thresholds are $0.25\,\mathrm{m}$ for root-height error, $0.8$ for the tilt feature, and $10\,\mathrm{m}$ for horizontal drift. The analytic soft versions use widths $0.02\,\mathrm{m}$, $0.04$, and $1\,\mathrm{m}$, respectively. Clip end is a timeout. Physical transitions rejected by the solver are flagged in replay.

\subsubsection{Task-specific rewards and settings}

\begin{figure*}[t]
    \centering
    \includegraphics[width=\textwidth]{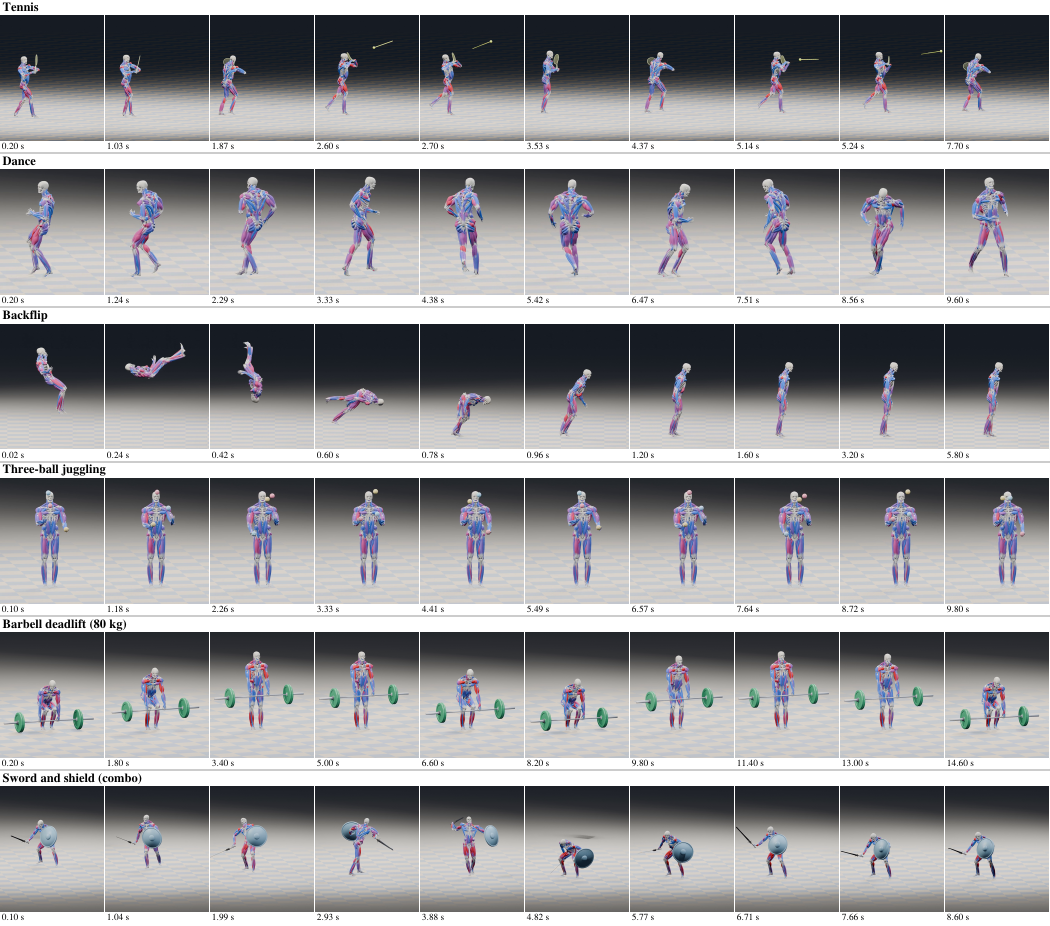}
    \caption{\textbf{Volumetric motion sequences.} Each row shows ten chronologically ordered snapshots from one recording; timestamps indicate elapsed video time. Rows show tennis, dance, backflip, three-ball juggling, barbell deadlift, and sword-and-shield motion. Muscle color indicates activation from blue to red.}
    \label{fig:supp-motion-sequences}
\end{figure*}

We use the tracking reward in Table~\ref{tab:supp-tracking-rewards} as the default. For better visualization, please refer to our accompanying video. Parentheses indicate the corresponding task's position in the video. Fig.~\ref{fig:supp-motion-sequences} shows sequences from the tennis, three-ball juggling, barbell deadlift, and sword-and-shield videos.

\paragraph{LaFAN1.}
LaFAN1 uses the default tracking reward without hand loads. ($0^{\prime}55^{\prime\prime}$)

\paragraph{Human video.}
We track the retargeted skeleton for walking and in-place exercise clips. ($2^{\prime}07^{\prime\prime}$)

\paragraph{Hand loads.}
We attach grasped props to the hands and incorporate their mass and inertia into the hand rigid bodies. This setup also applies to kettlebells and rackets. The world model, policy, and critic receive ten load features per hand: mass, three center-of-mass coordinates, and six independent inertia entries in the hand frame. The reference clip selects the same load in simulation and imagined rollouts.

\paragraph{Strength training.}
Lateral raises use a $3\,\mathrm{kg}$ dumbbell in each hand. Dumbbell deadlifts use $12$ or $22\,\mathrm{kg}$ per hand. Each dumbbell is fixed to the hand that holds it. Barbell deadlifts use total loads of $40$, $80$, $100$, and $140\,\mathrm{kg}$. In the $140\,\mathrm{kg}$ trial shown in the video, the character remains flexed at the hips and does not complete the reference lift. ($1^{\prime}20^{\prime\prime}$) We incorporate its mass and inertia into the right-hand body and keep its center of mass midway between the two grips. A weld constraint maintains the relative pose of the hands and transfers load between them. Kettlebell lifts use a single $12$ or $20\,\mathrm{kg}$ kettlebell held with both hands. ($1^{\prime}05^{\prime\prime}$)

\paragraph{AMASS box carrying.}
The carrying reference comes from the ACCAD subset of AMASS. The character holds a $2\,\mathrm{kg}$ box with both hands. We incorporate the box's mass and inertia into the right-hand body, preserving its center of mass, and weld the hands at the reference grasp. The horizontal root-tracking scale is $1.75\,\mathrm{m}$. ($1^{\prime}46^{\prime\prime}$)

\paragraph{Sword and shield.}
We retarget five Reallusion sword-and-shield clips distributed with ASE~\suppcite{peng2022ase}. All five clips use the same hand loads, with a $1.8\,\mathrm{kg}$ sword in the right hand and a $3.0\,\mathrm{kg}$ shield in the left. In the sword-tip tracking variant, the goal includes the reference positions of two points near the blade tip, and the reward adds a unit-weight penalty on their mean position error. The wrist-assisted variant adds three motor axes per wrist, with a torque limit of $12\,\mathrm{N\,m}$ per axis. ($1^{\prime}58^{\prime\prime}$)

\paragraph{Three-ball juggling.}
The juggling policy tracks the body motion of the cascade in Table~\ref{tab:supp-policy-architecture}. Each ball has a mass of $0.08\,\mathrm{kg}$. Within $40\,\mathrm{ms}$ of a scheduled catch, the script attaches the ball if its center is within one ball radius of the free receiving hand. ($1^{\prime}52^{\prime\prime}$)

\paragraph{Tennis.}
The policy tracks the reference swing from its first frame with a $0.32\,\mathrm{kg}$ racket fixed to the right hand. For the video, we resample the ball's initial position and velocity so that it reaches the racket's string bed at the intended contact time in the reference swing. The video shows two incoming balls and the corresponding returns. ($1^{\prime}38^{\prime\prime}$)

\section{Evaluation and Ablations}
\label{app:evaluation}

We measure tracking accuracy and success across motion clips. Controlled mechanical tests quantify how the mesh, solver settings, and tissue properties affect muscle loads and joint motion.

\subsection{Evaluation Protocols}

\subsubsection{Physical Evaluation Protocol}

For the LaFAN1 actuation-model comparison, we evaluate 47 references after excluding the \texttt{ground}, \texttt{fallAndGetUp}, and \texttt{pushAndFall} clips. We initialize one environment per clip at its first reference frame and run the frozen policy deterministically. Both backends use a $20\,\mathrm{ms}$ control interval, $1\,\mathrm{ms}$ rigid-body steps, and full-body ground contact. The volumetric backend updates tissue forces every $10\,\mathrm{ms}$.

We evaluate each clip for up to $10\,\mathrm{s}$ (500 control steps). A rollout is successful if it reaches $10\,\mathrm{s}$ or the end of the reference clip, whichever comes first, without triggering a failure due to root height, root tilt, root drift, a rejected solve, or a non-finite state. We retain simulated states up to the first termination for error analysis and exclude the terminating step and any later samples.

We report pelvis-relative body MPJPE by subtracting each trajectory's pelvis position and comparing the resulting offsets along the world axes. Root-local MPJPE expresses each offset in its own pelvis orientation frame before comparison. Both metrics exclude the pelvis itself and pool all valid control states across the 47 clips. We also record world-space MPJPE from the unaligned body positions. The normalized evaluation reward is
\begin{equation}
 \mathcal R_{\rm eval}=
 \frac{\sum_c\sum_{t\in\mathcal A_c}r_{c,t}}
      {r_{\max}\sum_c|\mathcal A_c|},
\end{equation}
where $\mathcal A_c$ contains the control steps for which clip $c$ is active at the start of the step, $r_{c,t}$ is the step reward, and $r_{\max}$ is the evaluator's reference reward per step. Evaluation throughput counts batched simulation transitions after the first five warm-up steps and includes policy inference, metric evaluation, and resets.

\subsubsection{Feedback and Learning Ablation Protocols}

The frozen-response baseline uses a fixed nominal generalized-load map $\widetilde Q^{\rm mus}(q,a)$ evaluated at the current skeletal pose $q$ and muscle activation $a$. In the closed-loop feedback ablation, we fit one MLP per muscle with two 64-unit SiLU hidden layers. The MLP predicts axial force from the distance between the tendon leads, the two signed tendon radii, and activation. We normalize the three lengths by the rest belly span. We reconstruct endpoint forces and routed tendon tensions using the original connection laws and map the reactions to skeletal loads through the attachment points and waypoints. Both variants solve the tissue equilibrium with doubled Young's modulus. Coupled applies the resulting loads, and Frozen applies the nominal map while using the same policy observations and skeletal model.

We fit the map on nominal-material solves from 187 calibration segments and independent per-muscle sweeps, with 60 separate segments for validation. All 247 segments are disjoint from the evaluation segments. The sweeps sample activation in $[0,1]$ and use per-muscle bounds on lead distances and routed radii derived from the skeletal kinematic tree with a $2\,\mathrm{mm}$ coordinate margin. The validation criterion is less than $1\%$ relative $\ell_2$ error for both attachment forces and joint torques. The nominal joint-torque error is $0.60\%$ on 790 validation poses and $4.76\%$ at the 1,000 held-out rollout initial states.

For the material perturbation, $\Delta_Q$ is $100$ times the $\ell_2$ norm of the difference between the frozen and coupled torques divided by the $\ell_2$ norm of the coupled torques. Both norms pool the 50 scalar joint torques across the same 790 matched pose and activation probes, excluding the free root. The error is $11.48\%$ with Young's modulus doubled, with a torque RMSE of $6.63\,\mathrm{N\,m}$.

The recorded-activation replay in Figure~\ref{fig:mechanical_response} uses a map that includes wrapping. For wrapped muscles, the inputs are the two lead positions in the cylinder frame, the two tendon radii, and activation. The map predicts endpoint forces, tendon tensions, contact forces at the leads, and the cylinder wrench. The wrapping input bounds come from the nominal training poses, padded by half a standard deviation per feature, with activation in $[0,1]$. Frozen applies the predicted loads directly during this replay. The curve ends when a query leaves the calibration domain, and independent fit validation for this map was not completed.

We hold policy weights fixed for feedback trials and use matched initial states and paired seeds. Activation/path probes isolate the mechanical response before closed-loop comparisons. Nominal proxy fit and perturbed response errors are reported separately. For learning-efficiency experiments, each algorithm/horizon is trained independently in the same volumetric dynamics. Shared-budget success and time-to-threshold are distinct statistics; failed threshold attainment is retained as censored data. Physical transitions exclude model predictions, and GPU-hours include model fitting and policy/value updates.

\subsection{Actuation and Learning Comparisons}

\subsubsection{Independent Line-Muscle and Volumetric Training}

We train a Hill-type line-muscle baseline\cite{zajac} and a volumetric controller independently with the same model-based learning procedure. Each run uses its own checkpoints and replay. For line-muscle lengths $\ell(q)$, moment arms $R(q)$ and tension $f_{\rm Hill}$, the baseline force map is

\begin{equation}
 Q_{\rm Hill}^{\rm mus}=R(q)^\top f_{\rm Hill}(q,v,a),
 \qquad \dot\ell=-R(q)v.
 \label{eq:hillforce}
\end{equation}

The two setups use the same skeleton, actuator correspondence, strength calibration, references, rewards, ground-contact settings, control rate, and common policy inputs. For the reported comparison, we use the same initial states and success criteria.

\subsubsection{Joint Loads at Matched States}
\label{sec:supp-same-state}

\begin{figure}[t]
  \centering
  \includegraphics[width=\linewidth]{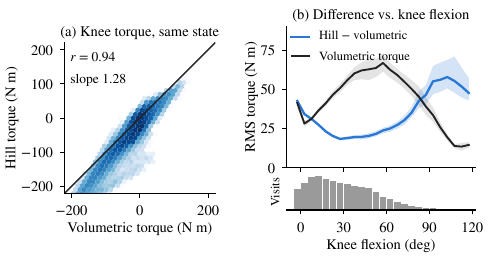}
  \caption{Knee load of the Hill line model versus the volumetric model at
  the same pose and activation, over states visited by the volumetric policy
  on LaFAN1 (both knees, 172k samples). (a) Density; line: identity.
  (b) RMS of the volumetric load (black) and of the Hill-minus-volumetric
  difference (blue) per 5$^\circ$ of flexion, with 95\% intervals from
  resampling clips; bottom: share of visited states.}
  \label{fig:supp-same-state}
\end{figure}

Table~\ref{tab:coupling} of the main paper compares controllers trained independently, each adapted to its own actuation, so it does not show how the two models differ at the same state. We run the volumetric policy closed loop on every LaFAN1 clip for up to 60\,s, restarting from the reference state after a fall, and record, at the first tissue solve of every other control step, the pose, the activation sent to the muscles, and the generalized muscle load of that solve. At each recorded state we evaluate the Hill load of Eq.~\ref{eq:hillforce} with the same activation. Both models use the same skeleton, actuator order, and anchor-to-joint force mapping, so the loads are compared in the same coordinates; the Hill passive load is a median 0.1\% of its total.

At the knee (Figure~\ref{fig:supp-same-state}) the loads agree in sign and trend ($r = 0.94$), but the Hill load is 28\% larger at equal activation, and the difference depends on pose: 18\,N\,m RMS at 25--30$^\circ$ of flexion against a 51\,N\,m load, 43 against 41\,N\,m below 0$^\circ$, and 48--58 against 13--34\,N\,m beyond 90$^\circ$. Table~\ref{tab:supp-same-state} lists the lower limbs and trunk. After the least-squares strength factor $s$ ($Q_{\mathrm{vol}} \approx s\,Q_{\mathrm{Hill}}$, per joint), the residual is still 32--41\% of the volumetric load at the hip ($x$) and knee joints and 48\% over all leg joints, so the difference is not a constant gain.

\begin{table}[t]
\centering\small
\caption{Hill versus volumetric generalized load at matched states, right /
left. RMS of the volumetric load; Pearson $r$; least-squares factor $s$ with
$Q_{\mathrm{vol}} \approx s\,Q_{\mathrm{Hill}}$; RMS residual after $s$,
relative to the volumetric RMS. Rotations about the model's $x$, $y$, $z$
joint axes.}
\label{tab:supp-same-state}
\setlength{\tabcolsep}{2.5pt}
\begin{tabular}{@{}lrrrr@{}}
\toprule
Joint & RMS (N\,m) & $r$ & $s$ & Residual \\
\midrule
Hip $x$   & 39.0 / 39.3 & 0.95 / 0.93 & 0.79 / 0.71 & 0.32 / 0.37 \\
Hip $z$   & 34.7 / 34.8 & 0.96 / 0.94 & 0.76 / 0.71 & 0.32 / 0.37 \\
Hip $y$   & 12.8 / 12.7 & 0.84 / 0.77 & 0.47 / 0.41 & 0.53 / 0.66 \\
Knee      & 50.0 / 47.7 & 0.93 / 0.92 & 0.71 / 0.65 & 0.36 / 0.41 \\
Ankle $x$ & 28.9 / 26.6 & 0.68 / 0.79 & 0.50 / 0.64 & 0.71 / 0.60 \\
Ankle $z$ & 8.7 / 10.2  & 0.24 / 0.34 & 0.06 / 0.18 & 0.98 / 0.95 \\
Ankle $y$ & 5.0 / 6.2   & 0.30 / 0.48 & 0.43 / 0.31 & 0.96 / 0.88 \\
\midrule
All leg joints & 111.2 & 0.88 & 0.65 & 0.48 \\
Spine and torso & 61.1 & 0.78 & 0.58 & 0.62 \\
\bottomrule
\end{tabular}
\end{table}

\subsubsection{Network Architecture Ablation}

We compare MLP and Transformer policies for motion tracking with the same 284-actuator wrapping model and world-model configuration. Both runs used 2,048 parallel environments across two GPUs, and 7,000 training iterations. We evaluated both checkpoints from the first frame of the two training clips, a LaFAN1 walking turn and a three-ball juggling motion, for up to $10\,\mathrm{s}$. The juggling score covers the humanoid motion. The MLP achieved lower tracking error on both clips, so we use it in the main experiments.

\begin{table}[t]
\centering\small
\caption{MLP and Transformer policy comparison on two training clips.
MPJPE uses pelvis-relative body positions in world axes and excludes the
pelvis. The overall MPJPE pools valid control states across both clips.}
\label{tab:supp-policy-architecture}
\begin{tabular}{@{}lrr@{}}
\toprule
Metric & MLP & Transformer \\
\midrule
Policy parameters & 8,354,620 & 7,627,676 \\
Training wall time (h) & 3.20 & 3.50 \\
Walking-turn MPJPE (mm) & 51.64 & 54.35 \\
Juggling MPJPE (mm) & 10.55 & 17.99 \\
Normalized return & 0.9113 & 0.9014 \\
\bottomrule
\end{tabular}
\end{table}

Both policies passed both clips (Table~\ref{tab:supp-policy-architecture}). The MLP reduced overall MPJPE by 14\% relative to the Transformer, with the larger reduction on the juggling clip. During the walking turn, mean horizontal pelvis errors were $2.53\,\mathrm{m}$ for the MLP and $2.69\,\mathrm{m}$ for the Transformer.

\subsubsection{Imagination Horizon}
\label{sec:supp-horizon}

The horizon $H$ sets how many steps actor gradients travel through the frozen world model. Let $D_k=\partial\hat s_k/\partial\theta$ be the sensitivity of the predicted state to the policy parameters, where the policy map $P$ includes both the actor and the goal construction. Applying the chain rule along the rollout gives
\begin{equation}
 D_{k+1}=(f_s+f_uP_s)D_k+f_uP_\theta,\qquad D_0=0,
\end{equation}
where $f_s$ and $f_u$ are the Jacobians of the world model with respect to state and action, and $P_s$ and $P_\theta$ are the Jacobians of the policy with respect to state and parameters. The recursion for $D_k$ multiplies Jacobians across rollout steps,
so errors in the world-model Jacobians can compound over a longer
actor horizon $H$ and affect the policy gradient.

\subsubsection{Running with Muscle Fatigue}

We adopt the 3CC-r fatigue model used in MuscleVAE~\cite{musclevae}. Each muscle has active, resting, and fatigued compartments. Their state limits the activation applied to the volumetric solver at each $20\,\mathrm{ms}$ control step and carries over across cycles of the $0.80\,\mathrm{s}$ LaFAN1 running reference. In a short test from the same initial pose, the frozen policy reached the $20\,\mathrm{s}$ limit with fatigue disabled. Scaling both fatigue and recovery rates by $25\times$ and $50\times$ reduced tracking time to $5.72$ and $3.56\,\mathrm{s}$, respectively, before root-height termination. Over the shared first $3.56\,\mathrm{s}$, mean local body-position error was $5.34$, $5.65$, and $9.89\,\mathrm{cm}$ for the three settings. All muscle solves were accepted in this short test.

\begin{figure}[t]
\centering
\includegraphics[width=\linewidth]{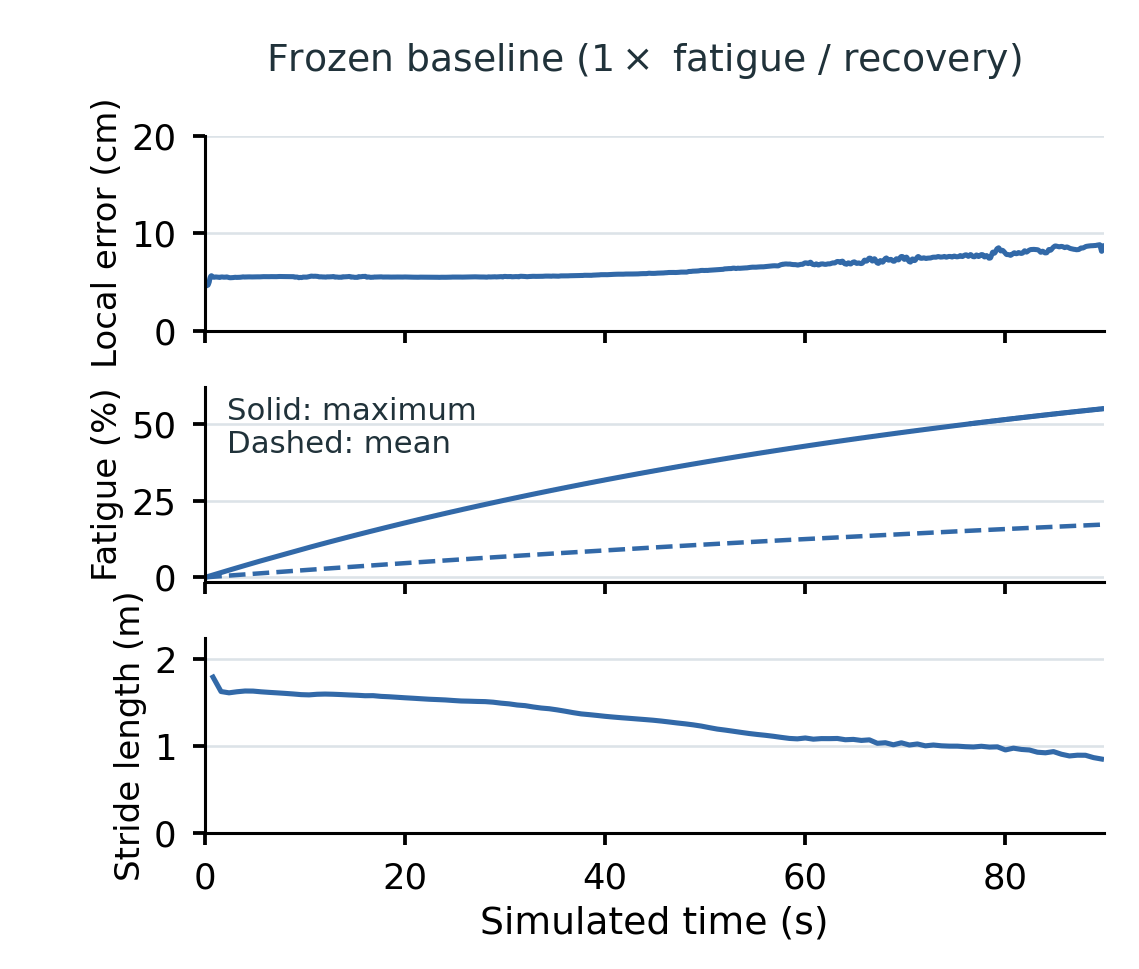}
\caption{\textbf{Long-horizon running with a frozen policy at fatigue and recovery rates ($1\times$).}
From top to bottom, the panels show local body-position error smoothed with a centered $0.80\,\mathrm{s}$ moving average, maximum and mean fatigue across muscles, and stride length estimated as forward pelvis displacement per complete $0.80\,\mathrm{s}$ reference cycle. All panels end at $89.82\,\mathrm{s}$, two seconds before the last accepted simulation step. Horizontal drift termination was disabled.}
\label{fig:supp-running-fatigue}
\end{figure}

\begin{figure*}[t]
    \centering
    \includegraphics[width=\textwidth]{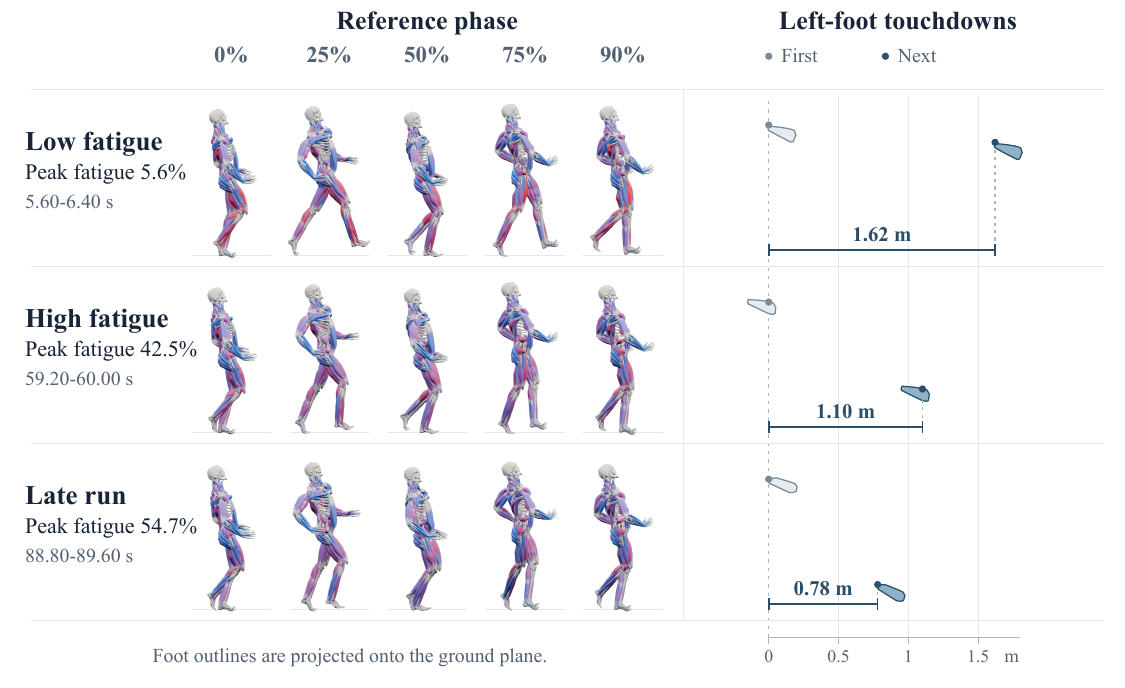}
    \caption{\textbf{Running gait across three fatigue stages.}
Rows show the three fatigue stages, and columns show snapshots at
0\%, 25\%, 50\%, 75\%, and 90\% of the reference gait cycle.
Peak fatigue is the largest fatigued fraction across the 284 muscle
units at the 25\% phase.
The right panel shows two consecutive left-foot touchdowns on a common
spatial scale, with the first touchdown aligned across rows.
Each dot marks the point where the foot first contacts the ground
at that touchdown.
The forward distance between the two dots gives the stride length,
which decreases from 1.62~m to 1.10~m and then to 0.78~m.}
\label{fig:fatigue-gait-touchdowns}
\end{figure*}

At fatigue and recovery rates, the frozen policy takes shorter strides as fatigue accumulates (Fig.~\ref{fig:supp-running-fatigue}). Across the three fatigue stages in Fig.~\ref{fig:fatigue-gait-touchdowns},
the forward distance between consecutive left-foot touchdowns decreases
from $1.62$ to $1.10$ and $0.78\,\mathrm{m}$.

We fine-tune the pretrained policy, world model, and critic with fatigue enabled during both simulator data collection and imagined rollouts. In a separate paired evaluation at nominal rates with horizontal drift termination disabled, the fine-tuned policy tracks for $124.58\,\mathrm{s}$, compared with $92.60\,\mathrm{s}$ for the frozen policy. Both show similar stride shortening. Measuring stride as forward pelvis displacement per reference cycle, the mean decreases from $1.64$ to $0.90\,\mathrm{m}$ for the frozen policy and from $1.73$ to $1.05\,\mathrm{m}$ for the fine-tuned policy. These means use the first and last ten complete cycles of the shared interval, truncated two seconds before the earlier termination.

\subsection{Numerical and Mechanical Ablations}

\subsubsection{Mesh-Resolution Ablation Protocol}
\label{app:mesh_ablation}

\begin{figure}[t]
\centering
\includegraphics[width=\linewidth]{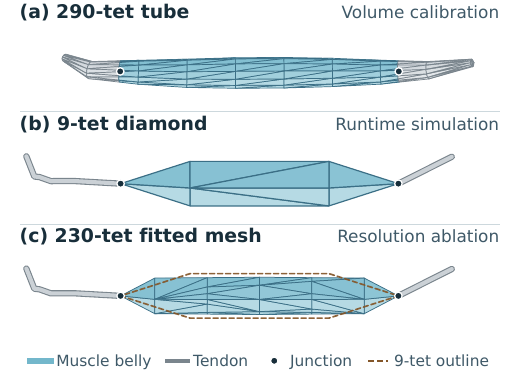}
\caption{\textbf{Three meshes of the left biceps brachii long head.}
Blue marks the central 64\% of the 290-tet tube's path and gray marks the two tendon regions. Gray lines in panels (b) and (c) trace the same routed tendons from the 9-tet and 230-tet belly tips to the bone attachments. The dashed line in (c) shows the 9-tet outline. Dots mark the two belly--tendon junctions. The views use the same scale and are aligned at the junctions.}
\label{fig:supp-muscle-mesh-variants}
\end{figure}

Figure~\ref{fig:supp-muscle-mesh-variants} compares the rest shapes of three meshes for the left biceps brachii long head. The 290-tet calibration tube follows the original muscle path and provides the target rest volume. The 9-tet runtime mesh represents the belly with two quadrilateral cross sections and two tips. The 230-tet fit rounds the coarse belly to approximate the calibration tube's profile, preserving the tip positions, rest volume, reference fiber directions, and tendon paths. Table~\ref{tab:mesh_ablation} compares the 9-tet and 230-tet simulation meshes, with 2,556 and 65,320 tetrahedra, respectively, across 284 muscle units. Both meshes use the same constitutive parameters, endpoint connections, tendon routing, wrapping cylinders, skeletal inertia, volume barrier, and stepping rates.

In the static comparison, we evaluate both meshes at the rest, walking, and running poses with activation $0.05$, and at the rest pose with activation $0.20$. At each matched state, $T_9$ and $T_{230}$ contain the routed tendon tensions. We report the relative discrepancy with respect to the coarse mesh,
\begin{equation}
 \Delta_T=100\,\frac{\|T_{230}-T_9\|_2}{\|T_9\|_2}.
\end{equation}
The largest discrepancy is $9.46\%$, at the rest pose with activation $0.20$.

For the dynamic comparison in Table~\ref{tab:mesh_ablation}, we start both meshes from the rest pose, hold activation at $0.05$, and simulate free motion for $1\,\mathrm{s}$. We compute the pooled joint-angle RMSE over the 50 completed control steps and 50 scalar joint coordinates, excluding the free root. The RMSE is $1.87^\circ$.

\paragraph{Muscle-driven elbow flexion.}
We apply the same $50\,\mathrm{Hz}$ activation sequence to the 9-tet and 230-tet models and integrate the elbow independently using the muscle forces computed by each model. Over $6\,\mathrm{s}$, the flexion-angle RMSE between the two trajectories is $2.37^\circ$.

The swept FEM model is built from the 290-tet tube mesh that serves as the basis for our visualizations. The tendons are passive regions of the same mesh. Their end rings follow the bones, and the resulting FEM reaction forces are applied to the skeleton to drive the elbow. After calibration against the 230-tet model, the swept model receives the same activation sequence and yields a flexion-angle RMSE of $9.77^\circ$ relative to the 230-tet trajectory.

\subsubsection{Solver Rate and Convergence Ablations}

Table~\ref{tab:solver_ablation} varies the muscle-equilibrium rate while holding the prescribed excitation update rate at $50\,\mathrm{Hz}$ and rigid integration at $1\,\mathrm{kHz}$. A second sweep varies the nodal residual limit from $1$ to $0.5$ and $0.25\,\mathrm{N}$ at a fixed muscle update rate of $100\,\mathrm{Hz}$. All trials use the same material and force laws, waypoint routing, volume barrier, excitation sequence, initial state, and hardware. A rejected solve terminates the trial.

\begin{table}[t]
\centering\small
\caption{\textbf{Numerical cost/accuracy ablation.} The reference uses
$1000\,\mathrm{Hz}$ muscle updates and a $0.001\,\mathrm{N}$ residual limit.
Throughput is the median over three runs in one FP64 CPU environment,
measured in accepted $20\,\mathrm{ms}$ control transitions per second.}
\label{tab:solver_ablation}
\begin{tabular}{@{}ccccc@{}}
\toprule
\shortstack{Muscle\\rate (Hz)} & \shortstack{Residual\\limit (N)} &
$\Delta_T$ (\%) & \shortstack{Work\\error (\%)} & Samples/s\\
\midrule
50 & 1 & 1.89 & 1.38 & 29.9 \\
100 & 1 & 0.82 & 0.53 & 22.2 \\
200 & 1 & 0.41 & 0.20 & 15.8 \\
100 & 0.5 & 0.87 & 0.56 & 19.1 \\
100 & 0.25 & 0.91 & 0.61 & 17.6 \\
\bottomrule
\end{tabular}
\end{table}

For $\Delta_T$, we evaluate all solver settings along the same reference elbow trajectory with the same prescribed activations. We compare the tension at both tendon ends on the $1\,\mathrm{ms}$ rigid-step grid. We compute the relative $\ell_2$ error over time and both ends for each muscle and average over muscles with nonzero reference tension.

In a separate rollout, we let the elbow move freely under the same activation sequence. We compare its angle and net muscle work with the reference. The work is $W=\sum_i Q_i^{\rm mus}(q_{i+1}-q_i)$, where $Q_i^{\rm mus}$ is the muscle torque applied during rigid substep $i$ and $q_i$ is the elbow angle in radians. We report $100\,|W-W_{\rm ref}|/|W_{\rm ref}|$ as the work error in percent. Each rollout lasts $2\,\mathrm{s}$, and the reference uses a nodal residual limit of $0.001\,\mathrm{N}$. We also record the number of rejected solves and the maximum nodal residual.

Table~\ref{tab:solver_ablation} uses a $1000\,\mathrm{Hz}$ reference. For the convergence test, all trials use $100\,\mathrm{Hz}$ muscle updates, including the reference, which uses a nodal residual limit of $0.001\,\mathrm{N}$. We compare tendon tensions along the same prescribed elbow trajectory with the same activations. To compare mechanical work, we let the elbow move freely in each run under the same activation sequence and sum the muscle work over that rollout. Relative to the $100\,\mathrm{Hz}$ reference, tightening the residual limit from $1$ to $0.25\,\mathrm{N}$ reduces the tension error from $0.266\%$ to $0.048\%$ and the work error from $0.101\%$ to $0.022\%$.

\subsubsection{Solver Comparison}

We benchmarked routed tetrahedral FEM with Newton iterations, the PD solver, and a quasi-static total-Lagrangian MPM implementation. Each world contains two biceps heads. The four inputs combine elbow coordinates of $-0.4$ and $-0.8\,\mathrm{rad}$ with activations of $0.1$ and $0.5$. Runs used an RTX 4070 SUPER with warm and cold starts and batches of one and 256 identical worlds. Timings cover the synchronized solver call with preloaded inputs and exclude routing and post-solve checks. For the 256-world warm-start batch, the median across the four per-input median latencies was $133.0$, $2.23$, and $113.6\,\mathrm{ms}$ for Newton, PD, and MPM, respectively. Each per-input median uses seven repeats.

\begin{figure*}[t]
    \centering
    \includegraphics[width=\textwidth]{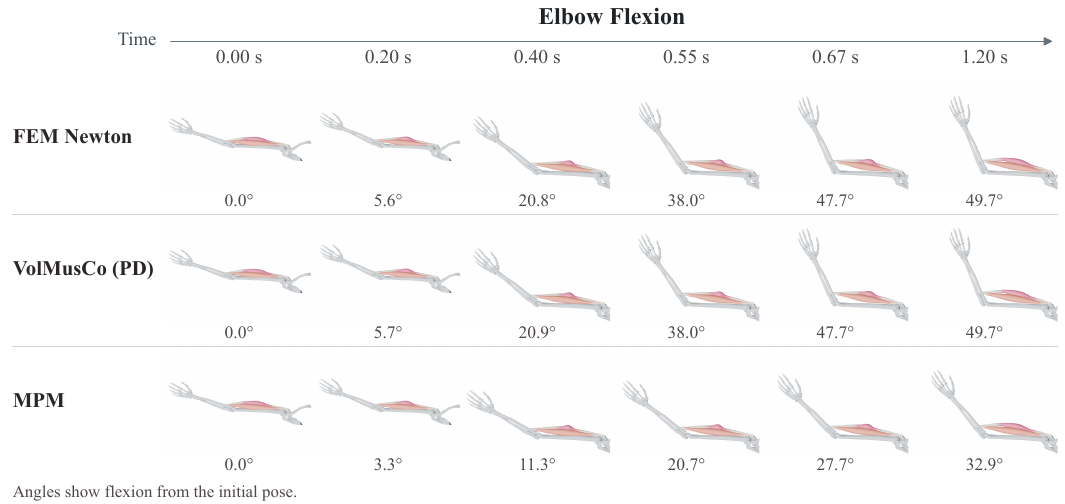}
    \caption{\textbf{Muscle-driven elbow flexion across solvers.}
    From left to right, the columns show elbow flexion at successive
    simulation times. The rows show FEM Newton, VolMusCo (PD), and MPM.
    All three solvers receive the same activation and advance the elbow
    using their own computed muscle forces. PD and FEM Newton produce
    nearly identical trajectories. MPM flexes more slowly and reaches
    a smaller flexion angle.}
    \label{fig:supp-solver-comparison}
\end{figure*}

All three methods met their respective $1\,\mathrm{N}$ stopping tests in these four cases. FEM and PD test tetrahedral nodal forces, and MPM tests its independent grid forces. Relative to the Newton solution, the maximum absolute elbow torque discrepancy in these cases was $0.020\,\mathrm{N\,m}$ for PD and $5.09\,\mathrm{N\,m}$ for MPM. Table~\ref{tab:supp-warm-solver-comparison} summarizes the warm-start
timings and torque errors. Fig.~\ref{fig:supp-solver-comparison} shows
the elbow motion produced by each solver.

\begin{table}[t]
\centering\small
\caption{\textbf{Warm-start solver comparison.} Each timed batch contains 256 identical two-biceps-head worlds. Times are medians across four cases, each timed seven times. The stopping threshold is $1\,\mathrm{N}$ on tetrahedral nodal forces for Newton and PD and on independent grid-node forces for MPM. Torque differences are maximum absolute elbow discrepancies relative to FEM Newton across the four cases.}
\label{tab:supp-warm-solver-comparison}
\begin{tabular}{@{}lccc@{}}
\toprule
Solver & \shortstack{Batch time\\(ms)} & \shortstack{Cases\\converged} & \shortstack{Torque diff.\\(N\,m)} \\
\midrule
FEM Newton & 133.0 & 4/4 & Reference \\
VolMusCo (ours) & 2.23 & 4/4 & 0.020 \\
MPM & 113.6 & 4/4 & 5.09 \\
\bottomrule
\end{tabular}
\end{table}

\subsubsection{Muscle Mass and Inertia}

The inertial prototype assigns mass to muscle vertices and adds the backward-Euler inertial term to the tissue objective. With mass matrix $M$, time step $h$, and predictor $y=x_n+h v_n+h^2g$, the term is $(x-y)^\top M(x-y)/(2h^2)$. The updated tissue shape transfers its tendon and contact reactions to the joint dynamics.

\begin{figure*}[t]
    \centering
    \includegraphics[width=\textwidth]{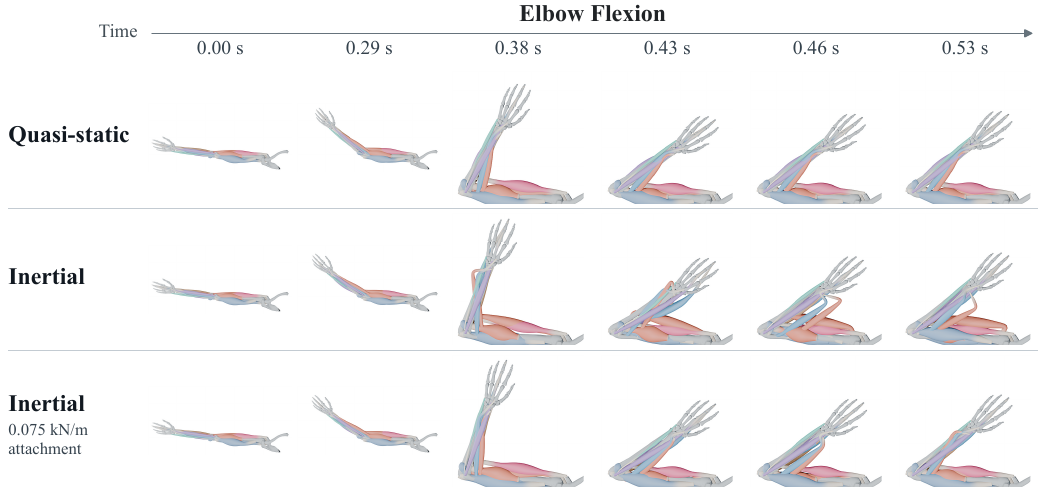}
    \caption{\textbf{Effect of muscle inertia during elbow flexion.}
    From left to right, the columns show elbow flexion at successive
    simulation times. The rows show the quasi-static model, the inertial
    model, and the inertial model with a $0.075\,\mathrm{kN/m}$ attachment
    spring. The compliant attachment reduces forearm muscle oscillations.}
    \label{fig:supp-elbow-inertia}
\end{figure*}

In the elbow trial, muscle inertia produced pronounced forearm muscle oscillations that compliant attachments reduced (Fig.~\ref{fig:supp-elbow-inertia}). In a full-body stress test, the inertial model took $76\,\mathrm{s}$ to simulate $1\,\mathrm{s}$ at $500\,\mathrm{Hz}$, and the coupled solve failed to  converge at the training update rates. We therefore use quasi-static  muscle equilibrium for training rollouts while retaining skeletal inertia.

\subsubsection{Local Fascial Coupling}

We repack the rest meshes of the medial and lateral gastrocnemius heads and the soleus into a single calf compartment. We define a common axis from the mean direction between the tendon tips and align the two cross-sectional rings of all three bellies within their overlapping axial interval. At each ring, three quadrilateral sectors tile a hexagonal outline. We size each sector to preserve the muscle's reference volume. For muscle $m$, we set the sector area $A_m$ using $V_m=A_m[L_s+(L_m-L_s)/3]$, where $L_s$ is the distance between rings and $L_m$ is the axial tip-to-tip span. The first term accounts for the central prism and the second for the two tapered ends. Each belly retains its ten independent vertices, nine tetrahedra, and original tendon endpoints and routing.

We pair the facing quadrilateral patches of neighboring bellies in the rest mesh. During the solve, we recompute the planes of each patch's two triangles from the deformed vertices and evaluate the signed normal gaps of the opposing vertices. We use the same unilateral penalty for these vertex--plane contacts and for contact between the belly-ring vertices and the sleeve walls:
\[
 E_{\mathrm{contact}}=\frac{1}{2}\sum_i k_i[-g_i]_+^2,
 \qquad [z]_+=\max(z,0),
\]
where $g_i$ is the signed clearance for sample $i$ and $k_i$ is its penalty stiffness. A penetrating vertex receives a normal force of magnitude $k_i[-g_i]_+$. At muscle--muscle interfaces, barycentric weights distribute the opposite force to the triangle vertices. The penalty acts along the normal, so the bellies can slide tangentially and separate.

The hexagonal fascial sleeve encloses the central compartment and follows the tibia. One scalar displacement $r$ allows the sleeve to expand or contract uniformly across its cross section while keeping its axial length fixed. Its tension-only membrane edges resist stretching, and a radial spring restores the rest cross section:
\[
 E_{\mathrm{sleeve}}(r)=
 \frac{1}{2}\sum_e k_e[\ell_e(r)-\ell_{e,0}]_+^2
 +\frac{1}{2}k_r r^2.
\]
Here $\ell_e(r)$ and $\ell_{e,0}$ are the current and rest lengths of edge $e$, $k_e$ is its stiffness, and $k_r=2500\,\mathrm{N/m}$. The sleeve limits outward bulging through normal contact and transmits the resulting wall reactions to the tibia.

\begin{figure*}[t]
    \centering
    \includegraphics[width=\textwidth]{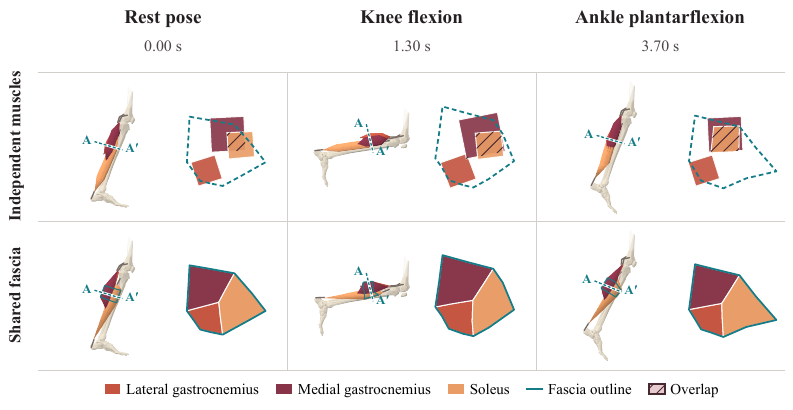}
    \caption{\textbf{Calf muscle deformation with and without shared fascia.}
From left to right, the columns show the rest pose, knee flexion,
and ankle plantarflexion. The top row shows independent muscle bellies,
and the bottom row shows muscle bellies in a shared fascial sleeve.
In the A--A$'$ sections, dashed outlines show the corresponding sleeve
boundary, and hatching marks overlap between muscle bellies.}
    \label{fig:supp-fascia-sliding}
\end{figure*}

At each tissue update, we solve for the muscle vertices and sleeve displacement together using the tissue, tendon, contact, and sleeve energies. In the knee--ankle demonstration (Fig.~\ref{fig:supp-fascia-sliding}), we update the tissue equilibrium at $30\,\mathrm{Hz}$ and hold the resulting tendon and sleeve loads for 16 rigid-body steps at $480\,\mathrm{Hz}$.

In a separate benchmark, we replayed 85 knee--ankle poses and muscle activation inputs for the three-muscle calf on one CPU thread in FP64. With the same packed meshes and material tangent correction, the mean warm-start time per tissue solve increased from $19.54\,\mathrm{ms}$ for independent bellies to $30.97\,\mathrm{ms}$ with sliding contact and the sleeve ($1.58\times$). The coupled solve updates all three bellies together. Fascial coupling could be incorporated into full-body training and evaluation in future work.

\begin{algorithm}[t]
\caption{Train a volumetric-muscle motion-tracking policy}
\label{alg:volumetric-training}
\small
\begin{algorithmic}[1]
\algrenewcommand{\algorithmicrequire}{\textbf{Input:}}
\algrenewcommand{\algorithmicensure}{\textbf{Output:}}
\Require Reference motions $\mathcal{D}$, muscle model $\mathcal{M}$
\Ensure Policy $\pi_\theta$, world model $W_\phi$, critic $V_\psi$
\Function{Train}{$\mathcal{D},\mathcal{M}$}
    \State $\mathcal{D}\gets$ \Call{RetargetToMASS284}{$\mathcal{D}$}
    \State Initialize $\pi_\theta,W_\phi,V_\psi$
    \State \Call{ResetToReference}{$\mathcal{D}$}
    \State $\mathcal{B}\gets$ \Call{CollectInitialTransitions}{}
    \For{each training iteration}
        \For{each parallel collection step}
            \State $(s,f)\gets$ \Call{ReadTrackingState}{}
            \State $o,g\gets$ \Call{ObservationAndGoal}{$s,\mathcal{D},f$}
            \State $a\gets$ \Call{BetaExplore}{$\pi_\theta(o,g)$}
            \State $s',b\gets$ \Call{Simulate}{$s,a,\mathcal{M}$}
            \State $d\gets$ \Call{TrackingDone}{$s',\mathcal{D},f+1$}
            \State $d\gets d\lor\neg b\lor$ \Call{ClipEnd}{$\mathcal{D},f+1$}
            \State \Call{Append}{$\mathcal{B},(s,a,s',f,d,b)$}
            \If{$d$}
                \State \Call{ResetToReference}{$\mathcal{D}$}
            \EndIf
        \EndFor
        \State \Call{TrainWorldModel}{$W_\phi,\mathcal{B}$}
        \If{world-model warm-up complete}
            \State \Call{TrainActorCritic}{$\pi_\theta,V_\psi,W_\phi,\mathcal{B},\mathcal{D}$}
        \EndIf
    \EndFor
    \State \Return $\pi_\theta,W_\phi,V_\psi$
\EndFunction

\Function{TrainWorldModel}{$W_\phi,\mathcal{B}$}
    \State $(s_0,a_0,s_1,\ldots,a_{K-1},s_K)\gets$ \Call{SampleWindow}{$\mathcal{B}$}
    \State $\hat{s}_0\gets s_0$
    \For{$t=0,\ldots,K-1$}
        \State $\hat{s}_{t+1}\gets$ \Call{WorldStep}{$W_\phi,\hat{s}_t,a_t$}
    \EndFor
    \State $\ell\gets$ \Call{StateContactLoss}{$\hat{s}_{1:K},s_{1:K}$}
    \State Mask invalid transitions and steps past episode end
    \State Update $\phi$ with the masked loss
\EndFunction

\Function{TrainActorCritic}{$\pi_\theta,V_\psi,W_\phi,\mathcal{B},\mathcal{D}$}
    \State $H_\pi\gets16,\quad H_V\gets32$
    \State $(s_0,f_0)\gets$ \Call{SampleValidStart}{$\mathcal{B}$}; $\hat{s}_0\gets s_0$
    \State Freeze $\phi,\psi$ for the actor update
    \For{$t=0,\ldots,H_\pi-1$}
        \State $o_t,g_t\gets$ \Call{ObservationAndGoal}{$\hat{s}_t,\mathcal{D},f_0+t$}
        \State $z_t\gets\mathrm{Enc}_\theta(g_t)$
        \State $a_t\gets\operatorname{sigmoid}(\mathrm{Dec}_\theta(o_t,z_t))$
        \State $\hat{s}_{t+1}\gets$ \Call{WorldStep}{$W_\phi,\hat{s}_t,a_t$}
        \State $r_t,\delta_t\gets$ \Call{ScoreStep}{$\hat{s}_{t+1},\mathcal{D}[f_0+t+1]$}
    \EndFor
    \State $J_\pi\gets$ \Call{SurvivalReturn}{$r,\delta,V_\psi,H_\pi$}
    \State Extend to $H_V$ with $\pi_\theta$ and gradients off
    \State Backpropagate $J_\pi$ through $H_\pi$ steps to update $\theta$
    \State $G^\lambda\gets$ \Call{LambdaReturns}{$r,\delta,V_\psi,H_V$}
    \State Fit $V_\psi$ to $G_t^\lambda$ for $0\le t<H_\pi$
\EndFunction
\end{algorithmic}
\end{algorithm}

\begin{algorithm}[t]
\caption{Step the volumetric musculoskeletal simulator}
\label{alg:volumetric-simulator}
\small
\begin{algorithmic}[1]
\algrenewcommand{\algorithmicrequire}{\textbf{Input:}}
\algrenewcommand{\algorithmicensure}{\textbf{Output:}}
\Require State $s$, muscle activations $a$, muscle model $\mathcal{M}$
\Ensure Returned state $s'$, step acceptance flag $b$
\Function{Simulate}{$s,a,\mathcal{M}$}
    \State $(q,v)\gets$ \Call{ReadSkeleton}{$s$}
    \State Hold $a$ fixed for the control step
    \For{each tissue update}
        \State \Call{UpdateAnchorsAndCylinders}{$q$}
        \For{each muscle $m$ in parallel}
            \State $x_m,b_m\gets$ \Call{Solve}{$x_m^-,q,a_m,\mathcal{M}_m$}
        \EndFor
        \For{wrapping muscles $m$ in parallel, if all $b_m$ are true}
            \State $x_m^0\gets$ \Call{WrapSeed}{$x_m,x_m^-,q,C_m$}
            \State $x_m,b_m\gets$ \Call{Solve}{$x_m^0,q,a_m,\mathcal{M}_m,C_m$}
        \EndFor
        \If{any $b_m$ is false}
            \State \Return \Call{ReadState}{$q,v,a$}, \textbf{false}
        \EndIf
        \State $x_m^-\gets x_m$ for all $m$
        \State $\{w_i\}\gets$ \Call{EndpointContactWrenches}{}
        \State $Q_{\mathrm{muscle}}\gets\sum_i J_i(q)^\mathsf{T}w_i$
        \For{each rigid substep in this interval}
            \State $q,v\gets$ \Call{RigidBodyStep}{$q,v,Q_{\mathrm{muscle}}$}
        \EndFor
    \EndFor
    \State \Return \Call{ReadState}{$q,v,a$}, \textbf{true}
\EndFunction

\Function{Solve}{$x,q,a,\mathcal{M},C=\emptyset$}
    \While{residual above tolerance and budget remains}
        \State $P\gets$ \Call{ProjectRotationAndVolume}{$x$}
        \State $F\gets$ \Call{NodalForces}{$x,q,a,\mathcal{M},C$}
        \State $x\gets$ \Call{VolumeSafePDStep}{$x,P,F$}
    \EndWhile
    \State \Return $x$, \Call{Accept}{$x,q,a,\mathcal{M},C$}
\EndFunction
\end{algorithmic}
\end{algorithm}

\clearpage

  \renewcommand{\refname}{Supplementary References}
  \renewcommand{\bibnumfmt}[1]{[S#1]}

\end{bibunit}

\end{document}